\documentclass{ws-ijmpa}
\usepackage[super,compress]{cite}
\usepackage{dcolumn}
\usepackage{bm}
\usepackage{xcolor}
\usepackage{graphicx}
\usepackage{adjustbox}

\usepackage[utf8]{inputenc} 
\usepackage[T1]{fontenc}    
\usepackage{hyperref}       
\usepackage{url}            
\usepackage{booktabs}       
\usepackage{nicefrac}       
\usepackage{microtype}      
\usepackage{mathtools}
\usepackage{commath}
\usepackage[sc,osf]{mathpazo}

\usepackage{float}
\usepackage{varioref}
\usepackage[sans]{dsfont}
\usepackage{tikz}
\usepackage{esint}

\makeatletter
\@for\next:={int,iint,iiint,iiiint,dotsint,oint,oiint,sqint,sqiint,
  ointctrclockwise,ointclockwise,varointclockwise,varointctrclockwise,
  fint,varoiint,landupint,landdownint}\do{%
    \expandafter\edef\csname\next\endcsname{%
      \noexpand\DOTSI
      \expandafter\noexpand\csname\next op\endcsname
      \noexpand\ilimits@
    }%
  }
\makeatother
\usepackage[makeroom]{cancel}
\DeclareMathOperator{\sign}{sign}
\newcommand{\Mathematica}{\textit{Mathematica\textsuperscript{\resizebox{!}{0.8ex}{\textregistered}}}}
\usepackage{xspace}
\def\8{\infty}
\def\oh{\frac{1}{2}}

\def\tt{\frac{2}{3}}

\def\ofi{\frac{1}{5}}

\def\d{\partial}
\def\i{\imath\,}

\def\dal{\partial_{\alpha}}
\def\dbe{\partial_{\beta}}
\def\dga{\partial_{\gamma}}

\def\undertext#1{\vtop{\hbox{#1}\kern 1pt \hrule}}
\def\ra{\rightarrow}
\def\Ra{\Rightarrow}

\def\VEV#1{\left\langle #1\right\rangle}

\def\tr{\hbox{tr}\,}
\def\diag#1{\hbox{diag}\left(#1\right)}
\def\dd#1{\frac{d}{d#1}}
\def\dbyd#1#2{\frac{d#1}{d#2}}
\def\pp#1{\frac{\partial}{\partial#1}}
\def\pbyp#1#2{\frac{\partial#1}{\partial#2}}

\def\bea{\begin{eqnarray} & &}
\def\eea{\end{eqnarray}}

\let\oldexp\exp
\renewcommand{\exp}[1]{\oldexp\left(#1\right)}

\def \R{\mbox{Re}}

\def\NS{Navier-Stokes}
\def\KH{Kelvin-Helmholtz}
\def\BS{Biot-Savart}

\def \BT{Burgers-Townsend}

\def\CVS{\textit{CVS}}

\def\val{v_{\alpha}}

\def\ral{r_{\alpha}}
\def\rbe{r_{\beta}}
\def\rga{r_{\gamma}}

\def\omu{\omega_{\mu}}
\def\onu{\omega_{\nu}}

\def\Xint#1{\mathchoice
   {\XXint\displaystyle\textstyle{#1}}%
   {\XXint\textstyle\scriptstyle{#1}}%
   {\XXint\scriptstyle\scriptscriptstyle{#1}}%
   {\XXint\scriptscriptstyle\scriptscriptstyle{#1}}%
   \!\int}
\def\XXint#1#2#3{{\setbox0=\hbox{$#1{#2#3}{\int}$}
     \vcenter{\hbox{$#2#3$}}\kern-.5\wd0}}

\def\Pint{\Xint/}

\newcommand{\Z}{\mathbb{Z}}

\DeclareMathOperator{\erf}{erf}
 
\renewcommand{\Re}{\textbf{Re }}
\renewcommand{\Im}{\textbf{Im }}

\begin{document}


\title{Confined Vortex Surface and Irreversibility. \\
2. Hyperbolic Sheets and Turbulent statistics}
\author{Alexander Migdal}
\address{Abu Dhabi Investment Authority,
211 Corniche Street, \\Abu Dhabi, United Arab Emirates}
\address{Department of Physics, New York University Abu Dhabi,
PO Box 129188, \\Saadiyat Island, Abu Dhabi, United Arab Emirates}

\maketitle
\begin{abstract}
 We continue the study of Confined Vortex Surfaces (\CVS{}) that we introduced in the previous paper.
 We classify the solutions of the \CVS{} equation and find the analytical formula for the velocity field for arbitrary background strain eigenvalues in the stable region.
 The vortex surface cross-section has the form of four symmetric hyperbolic sheets with a simple equation $|y| |x|^\mu =\mbox{const}$ in each quadrant of the tube cross-section ($x y $ plane).
 
 We use the dilute gas approximation for the vorticity structures in a turbulent flow, assuming their size is much smaller than the mean distance between them. We vindicate this assumption by the scaling laws for the surface shrinking to zero in the extreme turbulent limit.
 We introduce the Gaussian random background strain for each vortex surface as an accumulation of a large number of small random contributions coming from other surfaces far away. We compute this self-consistent background strain, relating the variance of the strain to the energy dissipation rate.
 
 We find a universal asymmetric distribution for energy dissipation.
 A new phenomenon is a probability distribution of the shape of the profile of the vortex tube in the $x y$ plane. 
 This phenomenon naturally leads to the "multifractal" scaling of the moments of velocity difference $v(\vec r_1) - \vec v(\vec r_2)$. 
 More precisely, these moments have a nontrivial dependence of $n, \log \Delta r$, approximating power laws with effective index $\zeta(n, \log \Delta r)$.
 We derive some general formulas for the moments containing multidimensional integrals. 
 The rough estimate of resulting moments shows the log-log derivative $\zeta(n, \log \Delta r)$ which is approximately linear in $n$ and slowly depends on $\log \Delta r$. However, the value of effective index is wrong, which leads us to conclude that some other solution of the \CVS{} equations must be found.
 We argue that the approximate phenomenological relations for these moments suggested in a recent paper by Sreenivasan and Yakhot are consistent with the \CVS{} theory. We reinterpret their renormalization parameter $\alpha\approx 0.95$ in the Bernoulli law $ p = - \frac{1}{2}\alpha \vec v^2$ as a probability to find no vortex surface at a random point in space.
 
\end{abstract}

\section{Introduction}

The ultimate goal of turbulence studies is to solve the \NS{} equations and determine why and how the solution covers some manifold rather than staying unique given initial data and boundary conditions. 

We also need to understand why it is irreversible even in the limit of zero viscosity when the \NS{} equations formally become the Euler equation corresponding to the reversible Hamiltonian system with conserved energy.

Once we know why and how the solution covers some manifold -- a degenerate fixed point of the \NS{} equations -- we would like to know the parameters and the invariant measure on this manifold.

The first obstacle to overcome on this path is to understand the irreversibility of the \NS{} dynamics in a limit when the viscosity goes to zero at fixed energy dissipation.

We know (or at least we assume) that the vortex structures in this extreme turbulent flow collapse into thin clusters in physical space. 
Snapshots of vorticity in numerical simulations  \cite{VTubes, Tubes19} show a collection of tube-like structures relatively sparsely distributed in space.

There was an excellent recent DNS \cite{S19} studying statistics of vorticity structures in isotropic turbulence with a high Reynolds number. The distribution of velocity circulation they have found was compatible with 2D vortex structures (instantons) suggested in \cite{M19d} following our early suggestions about area law for the velocity circulation \cite{M93}.

There were also some recent works modeling sparse vortex structures in classical \cite{VortexGasCirculation20}, and quantum \cite{QuantumCirculation21} turbulence.

We know such 2D structures in the Euler dynamics: these are vortex surfaces. Vorticity collapses into a thin boundary layer around the surface which is moving in a self-generated velocity field.
Such motion is known to be unstable against the \KH{} instabilities, which undermines the whole idea of random vortex surfaces.

However, the exact solutions of the \NS{} equations discovered in the previous century by Burgers and Townsend \cite{BURGERS1948, TW51} show stable planar sheets with Gaussian profile of the vorticity in the normal direction, peaked at the plane. 

Thus, the viscosity effects in certain cases suppress the \KH{} instabilities leading to stable, steady vortex structures.

This stable Gaussian solution is not the most general one; it takes some tuning of parameters in the background flow. This tuning is the central subject of our previous paper \cite{M21c} as well as this one.

Let us define the conditions and assumptions we take in this paper as well as the previous one.

We adopt Einstein's implied summation over  repeated Greek indexes $\alpha,\beta,\gamma,\dots$ , popular in theoretical physics, but not widely adopted in mathematical literature. We also use the conventional vector notation for coordinates $\vec r$, velocity $\vec v$ and vorticity $\vec \omega = \vec \nabla \times \vec v$.

\begin{enumerate}
    \item We address the turbulence problem from the first principle, the Navier-Stokes equation without external forcing.
    \item  We study an infinite isotropic flow in the extreme turbulent limit (viscosity going to zero at fixed anomalous dissipation rate). 
     \item The flow is assumed to be potential everywhere except narrow boundary layers surrounding closed vortex surfaces (shaped as tubes). 
    \item We restrict ourselves to the steady solutions (fixed points of Euler equations) and investigate the stability of these fixed points, leading to new boundary conditions of the Euler equations on vortex surfaces.
    \item As we shall argue, the effective internal forcing (spontaneous stochasticity) is generated by the infinite number of randomly located remote vortex structures, leading to a Gaussian random strain tensor.
    \item This Gaussian random matrix parametrizes our fixed point manifold, providing a calculable probability distribution.
\end{enumerate}

The recent research \cite{M21a, KS21} revealed that the \BT{} regime required certain restrictions on the eigenvalues and eigenvectors of the background strain for the planar vortex sheet solution.

This research is summarized and reviewed in our paper \cite{M21c}. The reader can find the exact solutions and the plots illustrating the vorticity leaks in the case of the non-degenerate background strain.

Abstracting from these observations, we conjectured in that paper that the local stability condition of the vortex surface with the local normal vector $\vec \sigma$ and local mean boundary value of the strain tensor $\hat S$ reduces to three equations (with $\vec v_\pm$ being the boundary values of velocity on each side)

\begin{eqnarray}\label{CVS}
&& \vec v_\pm \cdot \vec \sigma =0;\\
&&\hat S \cdot (\vec v_+- \vec v_-) =0;\\
&& S_{n n} =\vec \sigma \cdot \hat S \cdot \vec \sigma < 0
\end{eqnarray}

We call these equations the Confined Vortex Surface or \CVS{} equations. They are supposed to hold at every point of the vortex surface, thus imposing extra boundary conditions on the Euler equations. 

As the local strain tensor for the fixed vortex surface is uniquely determined by conventional Neumann boundary conditions for potential flow on each side, these additional \CVS{} boundary conditions restrict allowed shapes of vortex surfaces.

The first \CVS{} equation is simply a statement that the surface is stationary. Only the normal velocity must vanish for such a stationary surface-- the tangent flow around the surface reparametrizes its equation but does not move it in the Euler-Lagrange dynamics.

The second \CVS{} equation is an enhanced version of the vanishing eigenvalue requirement. The velocity gap $\Delta \vec v = \vec v_+- \vec v_-$ should be a null vector for this zero eigenvalue. This requirement is stronger than $\det \hat S  =0$.

The third equation demands that the strain pushes the fluid towards the surface on both sides.  

With negative normal strain and zero strain in the direction of velocity gap, the flow effortlessly slides along the surface on both sides, without leakage and pile-up.

Here is an intuitive explanation of the \CVS{} conditions-- they provide the permanent tangential flow around the surface, confining vorticity inside the boundary layer.

Once these requirements are satisfied in the local tangent plane, one could solve the \NS{} equation in the (flat) boundary layer and obtain the Gaussian \BT{} solution, vindicating the stability hypothesis. The error function of the normal coordinate will replace the velocity gap, and the delta function of tangent vorticity will become the Gaussian profile with viscous width.

The flow in the boundary layer surrounding the local tangent plane has the form \cite{MB16} (with $\Phi_\pm(r)$ being velocity potentials on both sides )
\begin{eqnarray}\label{BTSol}
&& \vec v\left(\vec r_0 + \vec \sigma_0 \zeta\right) = \oh (\vec v_+(\vec r_0) +\vec v_-(\vec r_0))+
\oh (\vec v_+(\vec r_0) -\vec v_-(\vec r_0))\erf{ \left(\frac{\zeta}{h \sqrt 2}\right)}; \\
&& \vec \sigma_0 = \vec \sigma(\vec r_0);\\
&& \vec v_\pm(\vec r_0) = \vec \nabla \Phi_\pm(\vec r_0);\\
&& \vec \omega\left(\vec r_0 + \vec \sigma_0 \zeta\right) = \frac{\sqrt  2}{h \sqrt \pi }\vec \sigma \times (\vec v_+(\vec r_0) -\vec v_-(\vec r_0))\exp{-\frac{\zeta^2}{2 h^2}};\\
&& h = \sqrt{\frac{\nu}{-\vec \sigma_0 \cdot \hat S(\vec r_0) \cdot \vec \sigma_0}};\\
&& \hat S_{\alpha\beta}(\vec r_0) = \oh \dal \dbe \Phi_+(\vec r_0) + \oh \dal \dbe \Phi_-(\vec r_0)
\end{eqnarray}
With the inequality $S_{n n} < 0$ satisfied, this width  $h$ will be real positive.

It is important to understand that this inequality breaks the reversibility of the Euler equation: the strain is an odd variable for time reflection, although it is even for space reflection.

Thus, we have found a dynamic mechanism of the irreversibility of the turbulent flow: the vorticity can collapse only to the surface with negative normal strain; otherwise, it is unstable.

\section{The exact analytic solution of the CVS equations}

In the first part of this study \cite{M21c} we have found some equations (\CVS{} equations) for the velocity field and this surface in cylindrical geometry.

The motivation for cylindrical geometry is two-fold. 
First, this is the only known case of exactly solvable stability equations. 
Second, several numerical simulations \cite{Tubes19, VTubes} indicate that the dominant vortex structures are long tubes corresponding to the cylindrical geometry.
There are also theoretical arguments we present below, justifying such solutions as self-consistent ones.

Still, there could exist more general 3D solutions for vortex surfaces surrounding finite volume.
Such solutions, if found, would be the best candidates to describe the turbulent statistics.

Let us describe, generalize and solve the cylindrical \CVS{} equations before studying the associated turbulent statistics. As we shall see, there is a simple analytic solution to this generalized equation, unlike the original one \cite{M21c}.

Before doing so, we need to make two more comments:

\begin{itemize}
    \item The cylindrical geometry does not mean the tube is a cylinder (the circle translated in the normal direction).
    It is another planar curve (hyperbolic or parabolic type) translated in the normal direction. There is no axial symmetry in our solution.
    \item We argue that hyperbolic solutions are the correct ones. The parabolic solutions are not unstable, as they satisfy all the CVS stability equations. We reject them because their velocity does not decrease at infinity as required by our boundary conditions.
\end{itemize}

\subsection{Stationary Vortex Sheets and Double-layer potentials}

The steady closed vortex surface $\mathcal S$ can be treated within the framework of hydrostatics, as it was recently advocated in our paper \cite{M21a}.

In the 3D space inside and outside the surface $ \mathcal S^\pm: \d \mathcal S^\pm = \mathcal S$ there is no vorticity so that the flow can be described by a potential $\Phi(\vec r)$ with the gap $ \Gamma = \Delta \Phi$ on the surface.

This is a well-known double-layer potential from electrostatics \cite{DL21}.
The potential is given by the Coulomb integral with dipole density $\Gamma$ plus a background constant strain potential
\begin{eqnarray}
   && \Phi_\pm(\vec r) = \oh W_{\alpha\beta} \ral \rbe -\frac{1}{4 \pi} \int_S  \Gamma_\pm(\vec r') d^2 \sigma_\alpha(\vec r') \dal \frac{1}{|\vec r - \vec r'|};\\
   && d^2 \sigma_\alpha(\vec r) = e_{\alpha\beta\gamma} d \rbe \wedge d\rga;
 \end{eqnarray}
 
The pressure in each of the domains inside/outside is given by the Bernoulli formula
\begin{eqnarray}
p_\pm = -\oh  \left(\dal  \Phi_\pm(\vec r)\right)^2
\end{eqnarray}

The normal velocity vanishes on both sides (Neumann boundary conditions for $\Phi_\pm$).

The tangent velocity, on the other hand, has a gap $\Delta \vec v$. This gap arises because of the gap in the potential
\begin{eqnarray}
&&\Gamma = \Phi_+(\vec r) - \Phi_-(\vec r); \; \forall \vec r \in \mathcal S;\\
&& \Delta \vec v = \vec \nabla \Gamma
\end{eqnarray}

If we take the simplest nontrivial solution for the inner potential $\Phi_-(\vec r)$ with constant strain
\begin{eqnarray}
&&\Phi_-(\vec r)  = \oh W^-_{\alpha\beta} \ral \rbe ;\\
&& W^-_{\alpha\alpha}  =0;
\end{eqnarray}
we arrive at nonlinear differential equation
\begin{eqnarray}\label{Req}
&&e_{\alpha\beta\gamma}\d_u R_\alpha(u,v) \d_v R_\beta(u,v)W^-_{\gamma\lambda} R_\lambda(u,v)=0;\\
&& \mathcal S: \vec r = \vec R(u,v);
\end{eqnarray}
This equation is equivalent to a linear equation
\begin{eqnarray}\label{LReq}
W^-_{\gamma\lambda} R_\lambda(u,v) = A \d_u R_\gamma(u,v) + B \d_v R_\gamma(u,v) 
\end{eqnarray}
with coefficients $A,B$ which can be reduced to constants by a choice of internal coordinates $u,v$ (2 d diffeomorphisms).

Thus, the equation can be solved by a superposition of exponential terms $\exp{-\alpha u - \beta v}$. However the the main \CVS{} equation for the vanishing projection of the strain on the velocity gap becomes a hard nonlinear problem.
\begin{eqnarray}
\left(\dal \dbe \Phi_+\left(\vec R(u,v)\right) + W^-_{\alpha\beta}\right) \left(\dbe \Phi_+\left(\vec R(u,v)\right) - W^-_{\beta\gamma} R_\gamma(u,v)\right) =0
\end{eqnarray}
It is not even clear that this problem has any solution in general geometry: by parameter counting, there are more equations than free parameters.

With cylindrical geometry 
\begin{eqnarray}
&&v = z;\\
&&\vec R(u,z) = \left(X(u), Y(u), z\right)
\end{eqnarray}
the \CVS{} equations can be solved in analytic form, thanks to the magic of complex analysis.

\subsection{CVS Equation for Cylindrical Geometry}

In the previous paper, \cite{M21c}, we considered a cylindrical geometry, where the piecewise harmonic potential has the following form:
\begin{eqnarray}\label{complexPot}
&&\Phi_\pm(x,y,z) = \oh a x^2 + \oh b y^2  + \oh c z^2 + \Re \phi_\pm(\eta);\\
&& a + b + c =0,\, a \le b \le c;\; c >0, a <0;\\
&& \eta = x+ \i y;\\
&& \phi_\pm(\eta) =  \frac{1}{2 \pi \i}\int d \Gamma_\pm(\theta) \log \left(\eta - C(\theta) \right);\\\label{Vgamma}
&&v_z = c z;\\
&& V_\pm(x,y) = v_x - \i v_y = a x - \i b y + F_\pm(x+ \i y);\\
&& F_\pm(\eta) = \phi'_\pm(\eta);
\end{eqnarray}

Here $C(\theta)$ is a complex periodic function of the angle $\theta \in (0, 2 \pi)$ parametrizing the closed loop $C$.
We simplified the notations compared to \cite{M21c}, now we denote the ordered eigenvalues of the background strain tensor as $a,b,c$.

\begin{figure}%
    \centering
      \includegraphics[width=\textwidth]{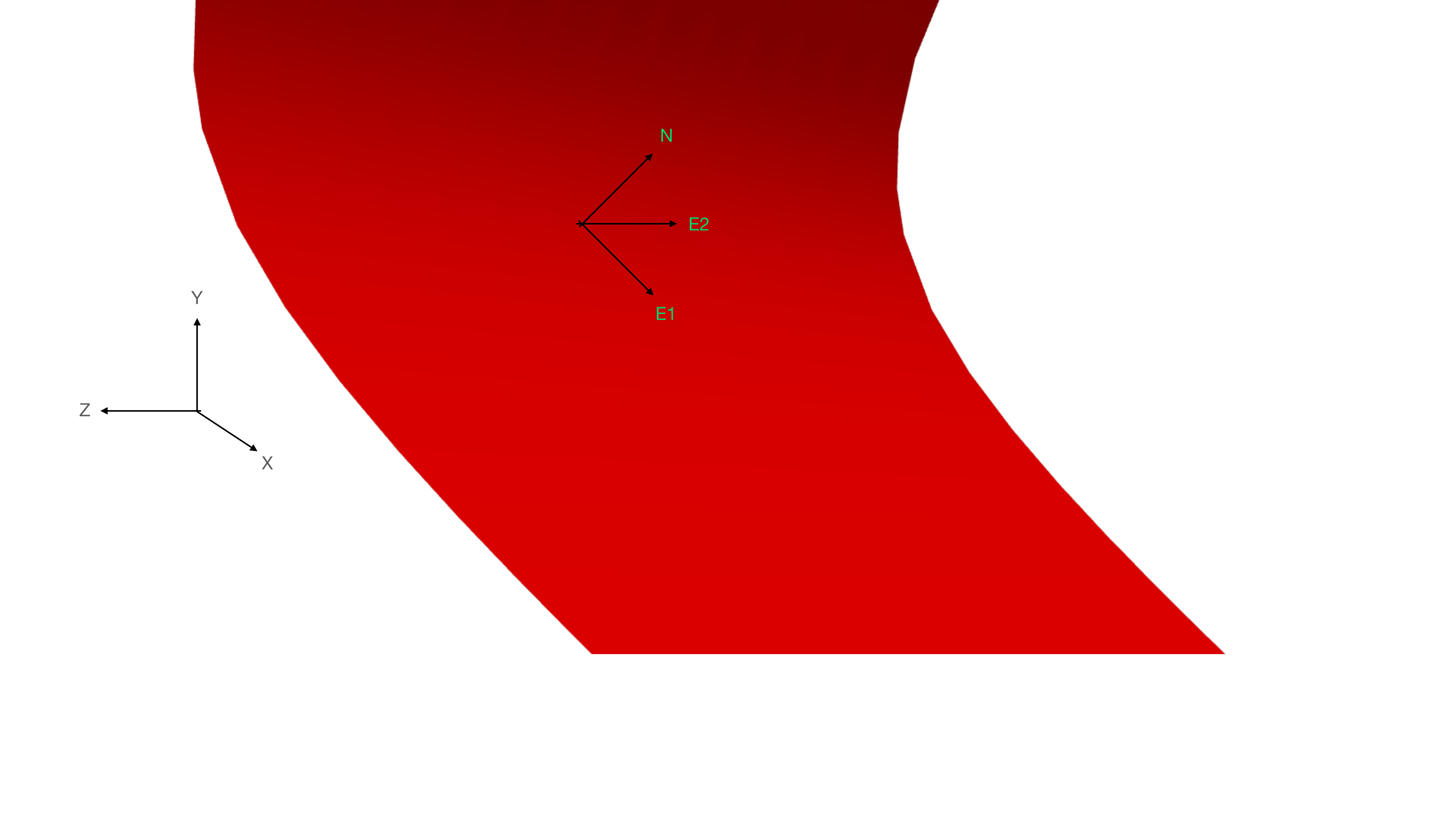}
    \caption{The  local tangent plane $E_1, E_2, N$ and the global Cartesian frame $X,Y,Z$.}
    \label{fig::TangentPlane}
\end{figure}

The geometry is as follows. The vortex surface corresponds to the parallel transport of the $x y $ loop $C$ in in the third dimension $z$. Thus, at some point on the surface, the local tangent frame is
\begin{eqnarray}
&&\vec E_1 =  \left(\frac{\Re C'}{|C'|}, \frac{\Im C'}{|C'|}, 0\right);\\
&&\vec E_2 =  \left(0,0, 1\right);\\
&& \vec E_3 = \left(\frac{\Im C'}{|C'|}, -\frac{\Re C'}{|C'|}, 0\right)
\end{eqnarray}
The first two orthogonal vectors define the tangent plane, and the third one $\vec E_3 = \vec E_1 \times \vec E_2$ corresponds to the normal $\vec \sigma$ to the surface. Do not confuse the $z$ direction with the normal -- this is one of the two tangent directions.
(See Fig.\ref{fig::TangentPlane}).

The functions $\Gamma_\pm(\theta)$ in this solution can be complex in general; the circulation will depend on the real part.

The double-layer potential studied in the previous paper \cite{M21c}, corresponds to the real $\Gamma_+(\theta)= \Gamma_-(\theta)$. In that particular case, the reparametrization of the curve would eliminate this function so that the solution would be parametrized by the loop $C(\theta)$ modulo diffeomorphisms.

As we see it now, the condition of real and equal $\Gamma_\pm(\theta)$ is unnecessary. We shall find simple analytic solutions by dropping this restriction.

The  condition of vanishing normal velocity at each side of the steady surface can be reduced to two real equations
\begin{eqnarray}\label{shapeEq}
&&\Im C'(\theta) V_\pm\left(C(\theta)\right) =0;
\end{eqnarray}

The second \CVS{} equation was reduced in \cite{M21c} to the complex equation (with $V_\pm$ denoting the boundary values at each side of the surface)
\begin{eqnarray}\label{CVSCyl}
V_+\left(C(\theta)\right) + V_-\left(C(\theta)\right) =0;
\end{eqnarray}

These two equations were derived in \cite{M21c} under assumptions of real $d\Gamma_\pm(\theta)$  in \eqref{Vgamma} which we find unnecessary and drop now.

In Appendix A, we re-derive these two equations without extra assumptions.

It follows from equation \eqref{CVSCyl} that one of the two eigenvalues of the local tangent strain at the surface vanishes. This vanishing eigenvalue corresponds to the velocity gap $\Delta V =V_+ - V_- = - 2 V_-$ as an eigenvector.

The way to prove this is to note that differentiation of \eqref{CVSCyl} by $\theta$ reduces to projecting the complex derivatives of $V_+(C)+ V_-(C)$  on the complex tangent vector $C'(\theta)$. As the $\theta$ derivative of $V_+(C)+ V_-(C)$ is zero in virtue of \eqref{CVSCyl}, so is the projection of the strain in the direction of the curve tangent vector $ C'(\theta)$. 

There is no normal velocity gap (as both normal velocities are zero), nor is there any gap in the $z$ component of velocity $v_z = c z$. Thus, the velocity gap aligns with the curve tangent vector $C'(\theta)$, and therefore the strain along the velocity gap vanishes. Direct calculation in Appendix A supports this argument.

The other eigenvalue $S_{n n}$, corresponding to the eigenvector $ \vec \sigma =\vec E_3$  normal to the surface, is then uniquely fixed by the condition of vanishing trace of the local strain tensor.

The third eigenvalue $c$  corresponds to the eigenvector $\vec E_2$ of our cylindrical tube.

Therefore, the normal component of the strain tensor is 
 \begin{eqnarray}
 S_{n n} = -c
 \end{eqnarray}
 
 \subsection{Spontaneous breaking of time-reversal symmetry}
 
 As the largest $c$ of the three ordered numbers $a,b,c $ with the sum equal to $0$ is always positive (unless all three are zero), we have a negative normal strain as required by the stability of the \NS{} equation in the local tangent plane.
 
 This condition does not restrict the background strain tensor, contrary to some of our previous statements. The irreversibility of turbulence manifests itself in the negative sign of the normal strain on the surface, which is true for arbitrary nonzero eigenvalues $a,b,c$. 
 
 The stability conditions rather restrict the vortex tube shape: its axis aligns with the leading eigenvector of the background strain, and its profile loop adjusts to the local strain to annihilate its tangent projection.
 
 The cylindrical shape of the vortex tube, unlike the general shape, guarantees the negative sign of the normal strain. Perhaps we are dealing with the spontaneous emergence of cylindrical symmetry at the expense of breaking the time-reversal symmetry from the stability condition. 
 
 The maximal value of the normal strain on the closed vortex surface represents a functional of its shape. So, the problem is to find under which conditions this maximal value is negative. The potential being harmonic, this normal strain equals minus the surface Laplacian of the boundary value of potential.
 
 Thus, we are dealing with the minimal value of the surface Laplacian of a boundary value of harmonic potential on a closed surface, with Neumann boundary conditions at this surface. Under which conditions is this minimal value positive? 
 If the answer is: only for the cylindrical surface, then the stability requires a cylindrical symmetry.
 
 This important problem deserves further study.
 
 The Euler equation is invariant under time reversal, changing the sign of the strain. Without the \CVS{} boundary conditions, both signs of the normal strain would satisfy the stationary Euler equation. Therefore, this \CVS{} vortex sheet represents a dynamical breaking of the time-reversal symmetry.
 
 Out of the two time-reflected solutions of the Euler equation, only the one with the negative normal strain survives. If virtually created as a metastable phase, the other one dissolves in the turbulent flow, but this remains stable. 
 
 Technically this instability displays itself in the lack of the real solutions of the steady \NS{} equation for positive $S_{n n}$. The Gaussian profile of vorticity as a function of normal coordinate formally becomes complex at positive $S_{n n}$, which means instability or decay in the time-dependent equation.
In \cite{KS21} the authors verified this decay/instability process. The time-dependent \NS{} equation was solved numerically in the vicinity of the steady solution with arbitrary background strain.
 Only the \BT{} solution corresponding to our \CVS{} conditions on the strain was stable.

 The breaking of time-reversal symmetry of the Euler dynamics is not driven by some external forces but rather is spontaneously created by internal \NS{} dynamics. The result of this microscopic stability mechanism of the \NS{} dynamics is the \CVS{} boundary conditions added to the ambiguous Euler dynamics of the vortex sheets.
 
 \subsection{Complex Curves}
 Let us now proceed with the exact solution of the \CVS{} equations.
 
 The \CVS{} equations for the cylindrical geometry reduce to the equation  \eqref{shapeEq} for the boundary loop $C(\theta)$ plus a complex equation $V_+(x,y) +V_-(x,y) =0$ at the loop $ (x,y) \in C$.
 
Consider two complex functions involved in this equation: $F_-(\eta)$ inside the loop, $F_+(\eta)$  outside.
These two functions are related through the boundary condition  
\begin{eqnarray}\label{Feq}
F_+(\eta) + F_-(\eta) + (a -b) \eta - c \bar \eta =0 ;\forall \eta \in C;
\end{eqnarray}

Let us look for the linear solution of the Laplace equation inside (constant strain):
\begin{eqnarray}
F_-(\eta) = (p + \i q) \eta
\end{eqnarray}
with some real $p,q$.

The vanishing  normal velocity from the inside leads to the differential equation for $C(\theta)  $. 
\begin{eqnarray}\label{formEq}
&&\Im \left( (2 p + a - b + 2\i q  ) C(\theta) -c \bar C(\theta)\right)C'(\theta) =0
\end{eqnarray}

This equation is the 2D version of the equations  \eqref{Req}, \eqref{LReq}.
It is integrable for arbitrary parameters. 
We shall use two dimensionless parameters $\gamma, \beta$  to parametrize $p,q$ as follows:
\begin{eqnarray}\label{params}
&&p + \i q =\frac{b-a}{2} +\frac{c e^{-2\i\beta }}{2 (2 \gamma +1) };
\end{eqnarray}

The general solution of \eqref{formEq} in polar coordinates reads \cite{MB14} (up to an arbitrary normalization constant)
\begin{eqnarray}\label{gensol}
&&C(\theta) =  e^{\i\beta}(1 + \i \tau) \tau^{\gamma};\\
&& \tau = \tan \theta
\end{eqnarray}
This solution applies when $\tau^\gamma$ is real.

It is a particular case of the general 3D solution \eqref{LReq}, outlined above.
The parameter $\tau = \exp{u}$ in the context of the general solution. 

\subsection{Parabolic curves}

Let us consider positive $\gamma$ first (the parabolic curves).

This solution passes through the origin, and it will have no singularity at $\tau=0$ in case $\gamma = n, n \in \Z, n \ge 0$. 

\begin{figure}%
    \centering
      \includegraphics[width=\textwidth]{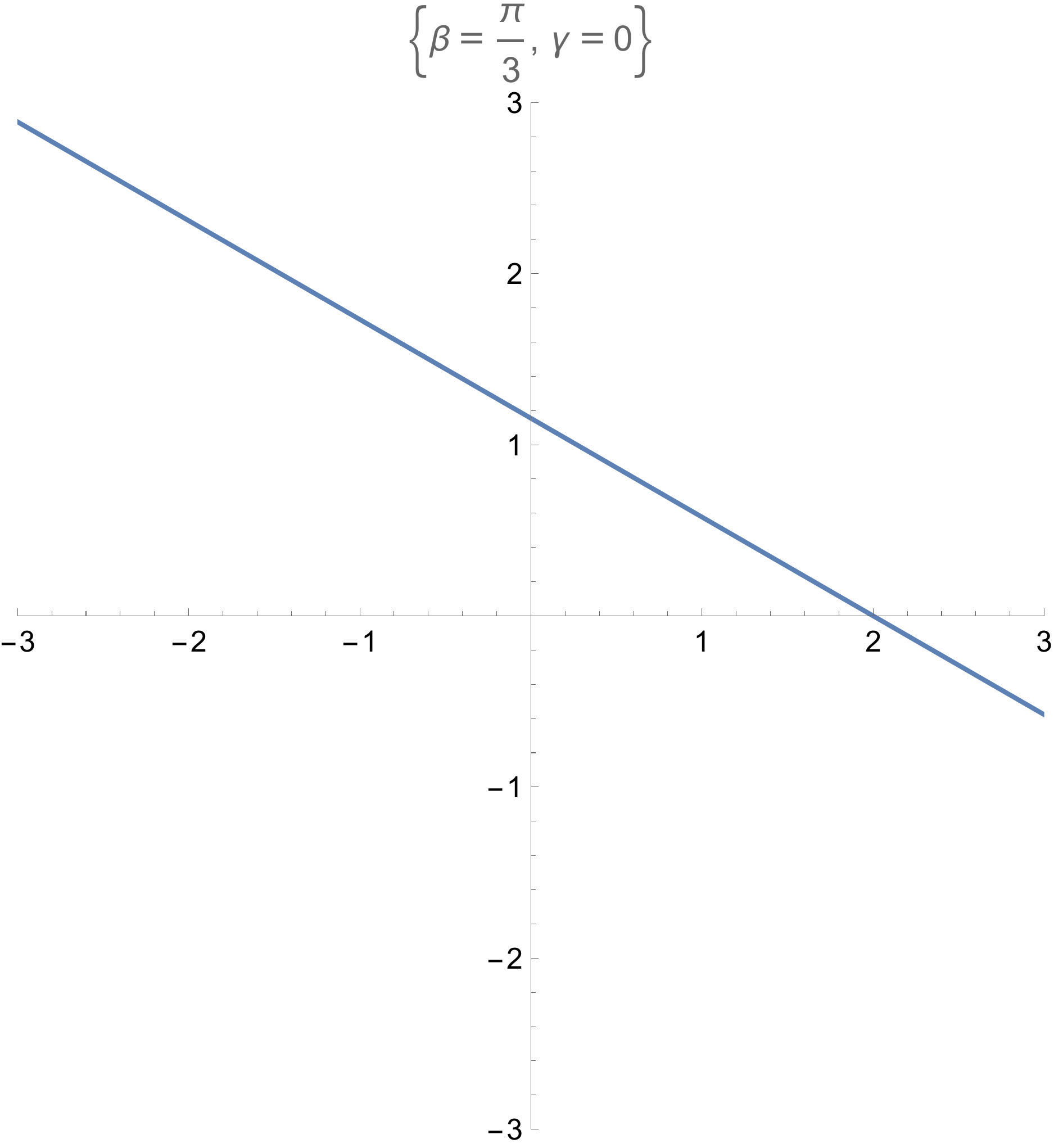}
    \caption{The  \BT{} case, $n=0$.}
    \label{fig::Curve0}
\end{figure}

\begin{figure}%
    \centering
      \includegraphics[width=\textwidth]{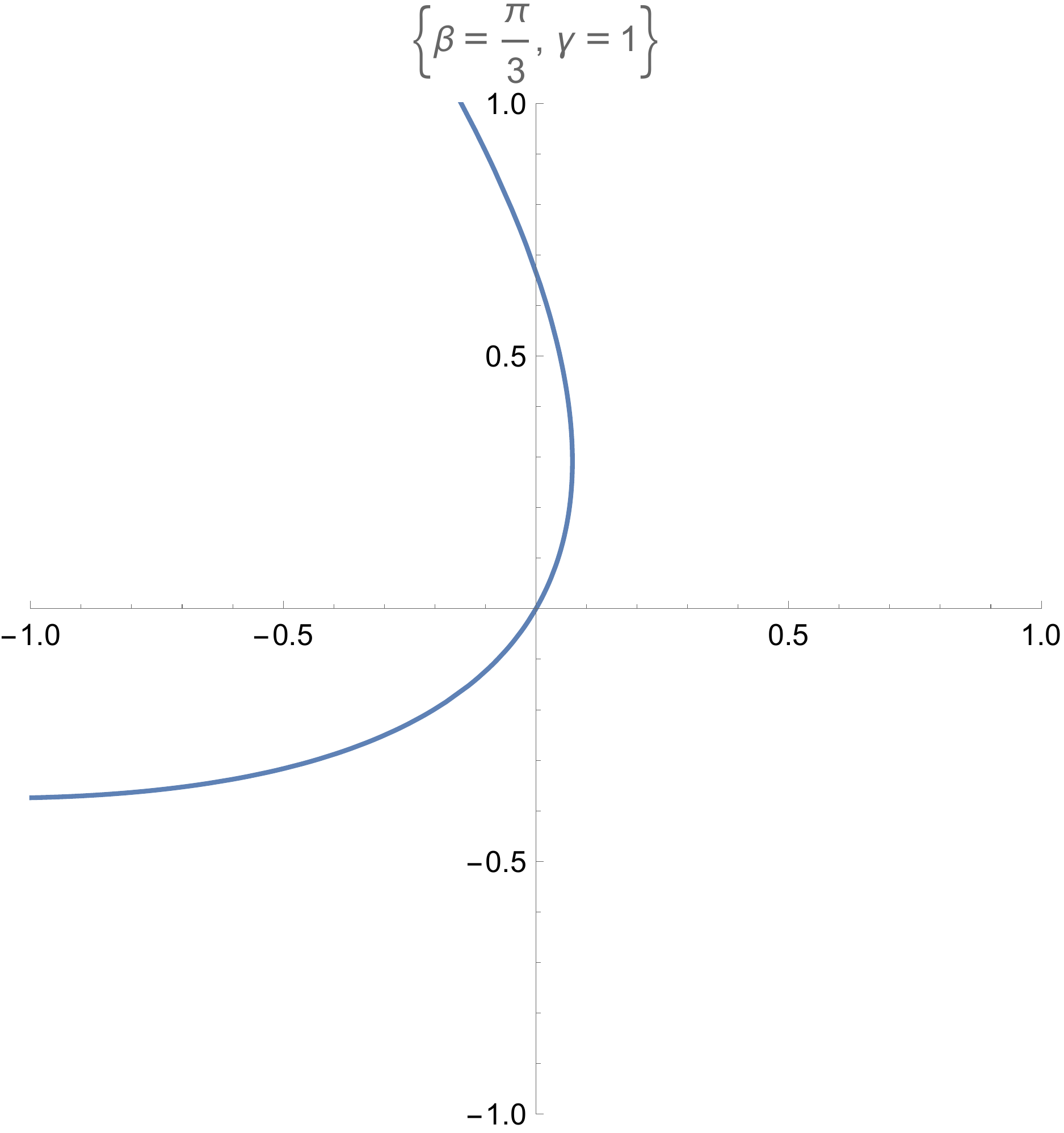}
    \caption{The \CVS{} for $n=1$.}
    \label{fig::Curve1}
\end{figure}

\begin{figure}%
    \centering
      \includegraphics[width=\textwidth]{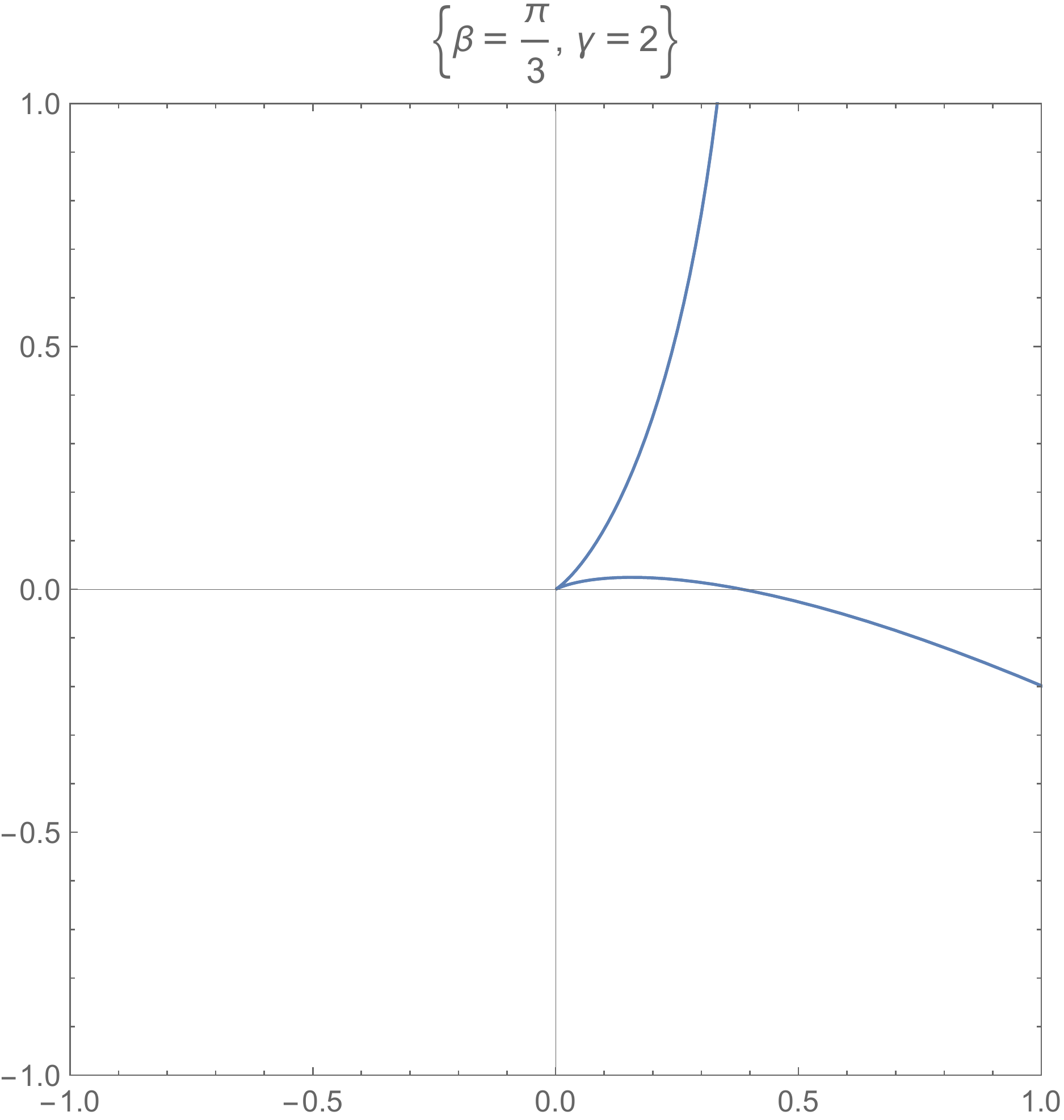}
    \caption{The \CVS{} for $n=2$.}
    \label{fig::Curve2}
\end{figure}

\begin{figure}%
    \centering
      \includegraphics[width=\textwidth]{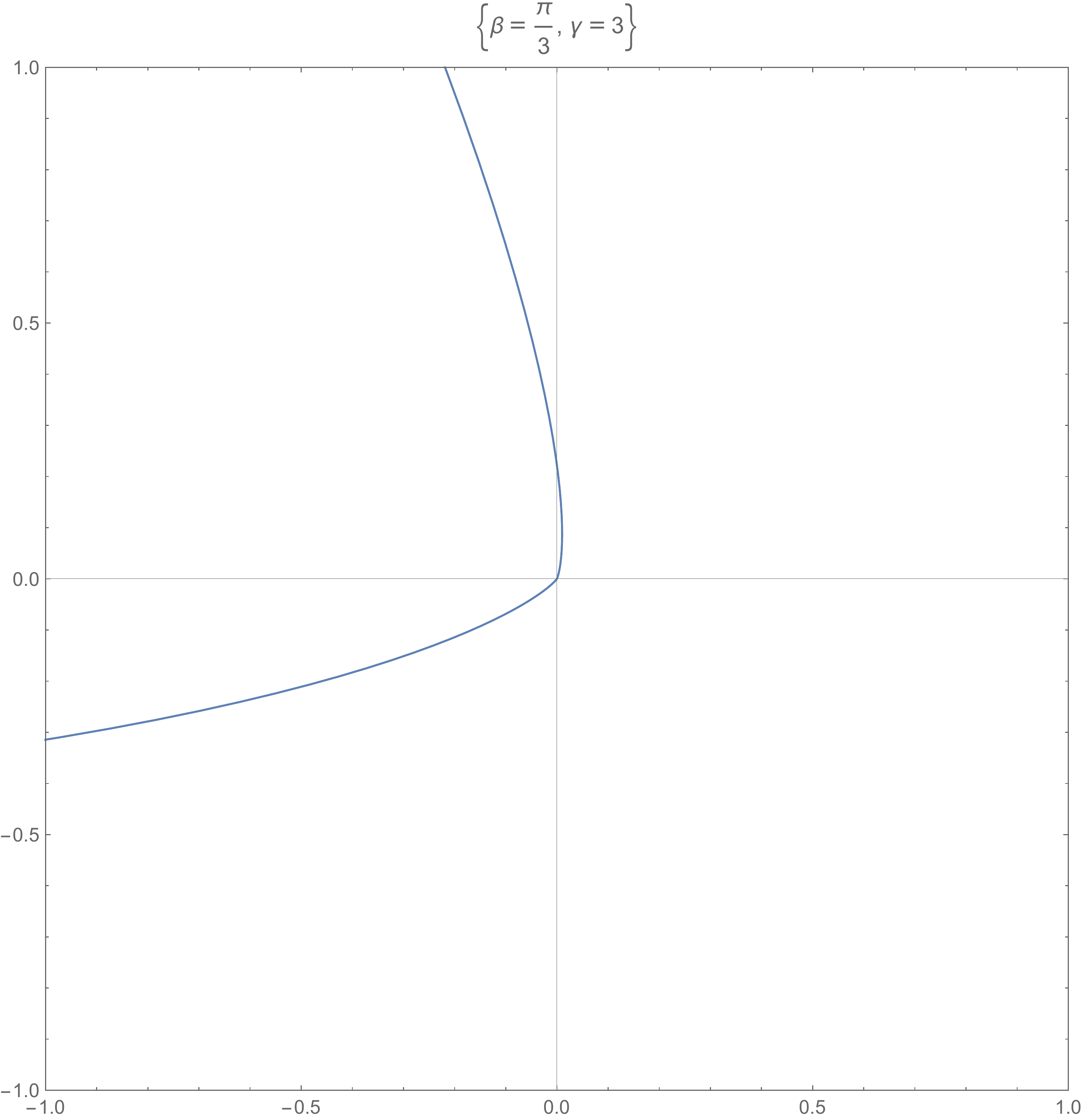}
    \caption{The \CVS{} for $n=3$.}
    \label{fig::Curve3}
\end{figure}

The case $n=0$ corresponds to a tilted straight line,  Fig.\ref{fig::Curve0}. This is just a \BT{} planar vortex sheet.
The next cases are already nontrivial Fig.\ref{fig::Curve1},\ref{fig::Curve2},\ref{fig::Curve3}.

However, not all positive $n$ are acceptable. For even positive $n$ we have a cusp in a curve, because only positive $r(\theta)$ exist at $\theta\ra 0$. This is clearly visible at $n=2$ (Fig.\ref{fig::Curve2}).

For odd positive $n= 2 k+1$ there is a linear relation between $x,y$ at small $\tau$ provided $\beta \neq 0$.
The slope of the curve $y(x)$  at $x= \pm 0$ is the same.

Higher derivatives are singular, as we have 
\begin{eqnarray}
&&x \sim \tau^{2 k +1}  \cos \left(\beta\right) (1 - \tan\left(\beta\right) \tau) ;\\
&&y \sim \tau^{2 k +1}  \sin \left(\beta\right) (1 + \cot\left(\beta\right) \tau) ;\\
&& y \ra \tan \left(\beta\right) x \left( 1 + \frac{2\sign x }{\sin\beta} \left(\frac{|x|}{\cos \left(\beta\right)}\right)^{\frac{1}{2 k +1}} \right);\\
&& \dbyd{y}{x} \ra \tan \left(\beta\right) \text{ at } x \ra \pm 0
\end{eqnarray}

As the slopes $\dbyd{y}{x}$  are the same at $x = \pm 0$, the vortex sheet reduces to a plane up to higher-order terms. Thus, the \BT{} solution, assuming the infinitesimal thickness of the vorticity layer, will still apply here. Ergo, the sheet will be stable at this point and the rest of the surface.

Thus, we have found the discrete parametric family of \CVS{} shapes. It simplifies in terms of $\tau = \tan\theta$.
\begin{eqnarray}\label{Acurve}
&&\eta(\tau) = e^{\i \beta} \left(1 + \i \tau\right) \tau^{2 k + 1};\\
&&k \in \Z, k \ge 0;\\
&& -\infty < \tau < \infty;
\end{eqnarray}

We have a conformal map from the $\tau $ plane to the plane of $\eta = x + \i y$, with our curve $C$ corresponding to the real axis. The physical region outside the loop corresponds to the sector in the lower semiplane, $ -\frac{ \pi}{( k +1)}< \arg \tau < 0$.

The infinity in the physical space corresponds to infinity in the $\tau$ plane.

There are two singular points corresponding to the vanishing derivative $\eta'(\tau) =0$. This happens at
\begin{eqnarray}
&&\tau = 0;\\
&& \tau =\frac{\i (2 k+1)}{2 (k+1)};
\end{eqnarray}

The first one maps to the origin in the $\eta$ plane.  The second one is in the upper $\tau$ semiplane, so it does not affect the inverse function $\tau(\eta)$ in the physical sector of the lower semiplane.

The next step is to find the holomorphic function $F_+(\eta)$ which defines the complex velocity on the external side of the sheet (the internal side has linear velocity $F_-(\eta) = (p + \i q) \eta$ ).

From the \CVS{} equation we get the boundary values, which we express in terms of $\eta, \eta^*$
\begin{eqnarray}\label{boundaryF}
 &&F_+(\eta) = -(p + a - b+ \i q)\eta  + c \eta^* ;\; \forall \eta \in C
\end{eqnarray}
We have to continue this function from the curve \eqref{Acurve} into the part of the complex plane outside of this curve.

Let us express the right side in terms of $\tau$
\begin{eqnarray}\label{param}
&&f(\tau) = -(p + a - b+ \i q)\eta(\tau)  + c \tilde \eta(\tau)
\end{eqnarray}
Here $\tilde \eta(\tau)$ is the complex conjugate expression for $\eta$
\begin{eqnarray}
\tilde\eta(\tau) = e^{-\i \beta} \left(1 - \i \tau\right) \tau^{2 k + 1};\\
\end{eqnarray}
At the real axis of $\tau$ the variables $\eta(\tau),\tilde\eta(\tau)$ will be complex conjugates, and the boundary value
\begin{eqnarray}
f(\tau) = F_+(\eta(\tau)); \forall \Im \tau =0
\end{eqnarray}

Now we have an obvious analytic continuation to the lower semiplane of $\tau$. After some algebra we get \cite{MB14}
\begin{eqnarray}
&&\eta(\tau) = e^{\i \beta} \left(1 + \i \tau\right) \tau^{2 k + 1};\\
&&f(\tau) = \frac{1}{2} e^{-i \beta } \tau ^{2 k+1}\nonumber\\
&&\left(\frac{c (-i (8 k+7) \tau +8 k+5)}{4 k+3}-i e^{2 i \beta } (\tau -i) (2 a+c)\right)
\end{eqnarray}

This parametric function in the sector $ -\frac{ \pi}{( k +1)}< \arg \tau < 0$ of the complex $\tau$ plane represents an implicit solution of the \CVS{} equation.

Now, let us consider the limit $\tau \ra \infty$ corresponding  to $\eta \ra \infty$. In this limit, the function $F_+(\eta(\tau))$ linearly grows 
\begin{eqnarray}
&& F_+(\eta) \ra  B \eta;\\
&& B =-a+\frac{1}{2} c \left(-1-\frac{e^{-2 i \beta } (8 k+7)}{4 k+3}\right)
\end{eqnarray}

This means that the true value of the $x y$ components of strain at infinity is not  $\diag{a,-c-a}$ we thought it was. It is instead
\begin{eqnarray}
\hat S = \left(
\begin{array}{cc}
 -\frac{c ((8 k+7) \cos (2 \beta )+4 k+3)}{8 k+6} & -\frac{c (8 k+7) \sin (2 \beta )}{8 k+6} \\
 -\frac{c (8 k+7) \sin (2 \beta )}{8 k+6} & \frac{c ((8 k+7) \cos (2 \beta )-4 k-3)}{8 k+6} \\
\end{array}
\right)
\end{eqnarray}

The eigenvalues of this strain are 
\begin{eqnarray}
\left\{-\frac{c (6 k+5)}{4 k+3},\frac{2 c (k+1)}{4 k+3}\right\}
\end{eqnarray}

The angle $\beta $ dropped from the eigenvalues because it can be eliminated by the $U(1)$ transformation $\eta \ra \eta e^{-\i \beta}$, which leaves the eigenvalues invariant.

The parabolic solution does not decrease at infinity, so it is inadequate for the turbulent statistics we develop later in this paper.

\subsection{Hyperbolic curves}

Let us now consider the hyperbolic curves with negative $\gamma< 0$, which do not need to be half-integer. 
We redefine our parameters as follows:
\begin{eqnarray}
&&\eta(\xi) = \pm e^{\i \beta} \left(\xi +\i\xi^{-\mu}\right);\\
&& \xi = \tau^{\gamma};\\
&& \mu = -1 -\frac{1}{\gamma }; \mu >0;\\
&& 0 < \xi < \infty
\end{eqnarray}

\begin{figure}%
    \centering
      \includegraphics[width=\textwidth]{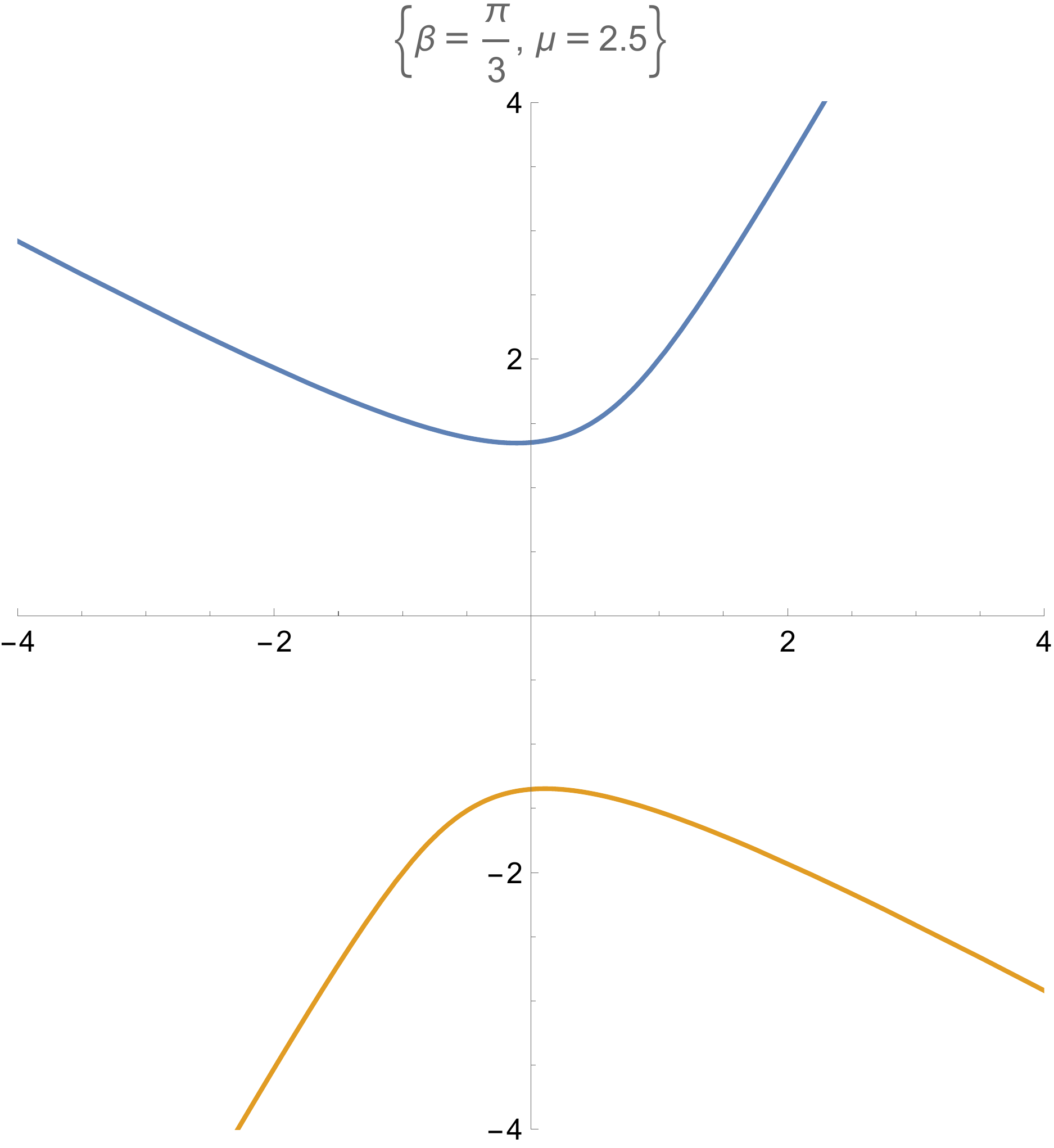}
    \caption{The hyperbolic \CVS{} for $\beta = \pi/3, \mu = 2.5$.}
    \label{fig::Hyperbola1}
\end{figure}

\begin{figure}%
    \centering
      \includegraphics[width=\textwidth]{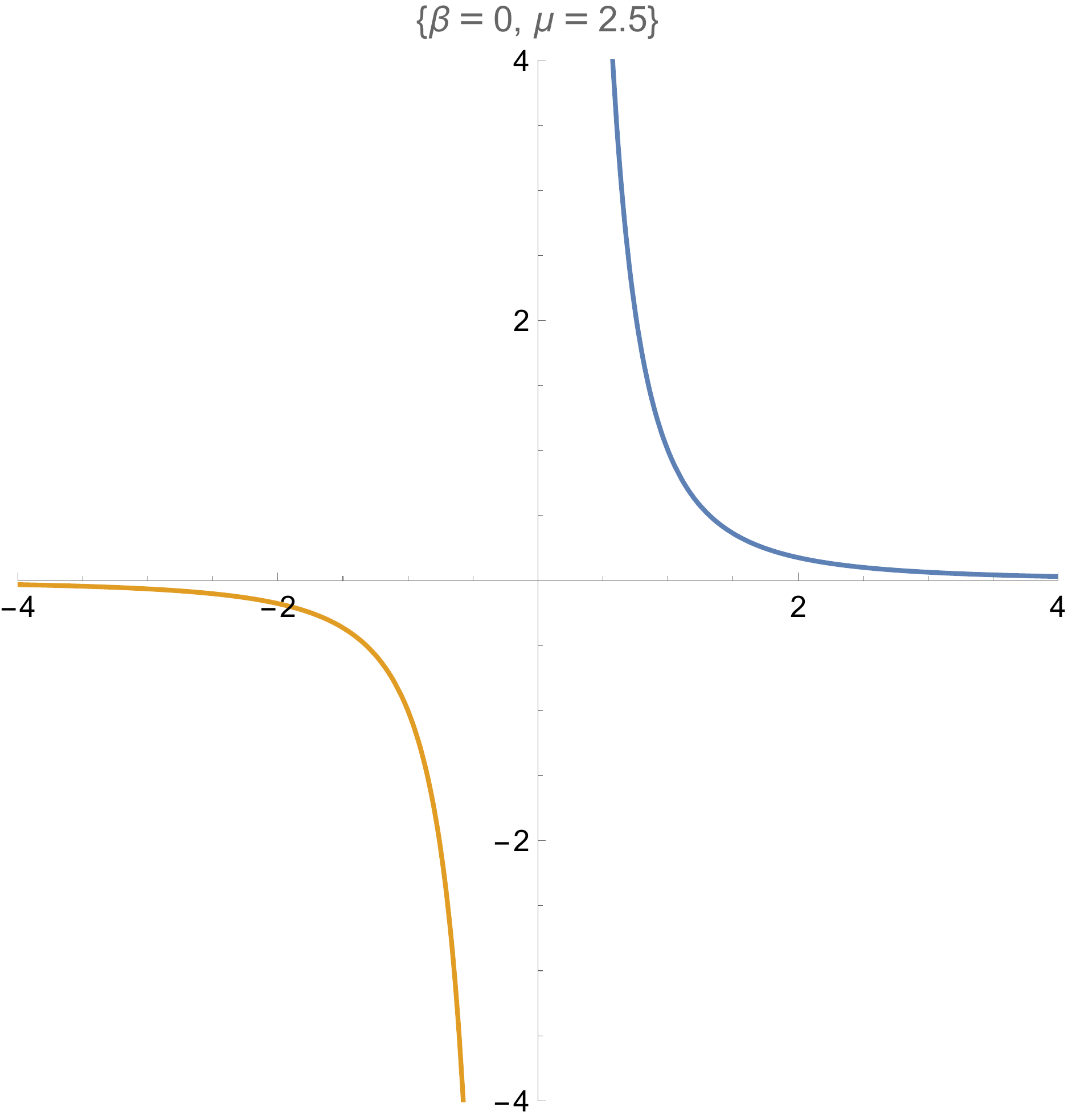}
    \caption{The hyperbolic \CVS{} for $\beta = 0, \mu = 2.5$}
    \label{fig::Hyperbola2}
\end{figure}

The two curves on Fig. \ref{fig::Hyperbola1}, \ref{fig::Hyperbola2} correspond to various $\mu, \beta$.
Note that there are no singularities at the origin, as both curves are going around it.

There is a singularity of the inverse function at the point in the upper semiplane:
where $\eta'(\xi_0)=0$:
\begin{eqnarray}
\xi_0 = (i \mu )^{\frac{1}{\mu +1}}
\end{eqnarray}
Therefore, we need to use the lower semiplane as a physical domain for $f(\xi ) =F_+(\eta(\xi))$.

The holomorphic function $f(\xi)$ is reconstructed by means of the same steps as in the parabolic case. We get
\begin{eqnarray}
&&f(\xi) =\frac{1}{2} e^{-i \beta } \xi  \left(\frac{c (\mu -3)}{\mu -1}-e^{2 i \beta } (2 a+c)\right)-\nonumber\\
&&\frac{i e^{-i \beta } \xi ^{-\mu } \left(c (3 \mu -1)+e^{2 i \beta } (\mu -1) (2 a+c)\right)}{2 (\mu -1)}
\end{eqnarray}

Let us now find the limit at $\xi\ra \infty$, when $\eta$ and $f$ both linearly grow
\begin{eqnarray}
&&f \ra \eta B;\\
&& B = -a+\frac{1}{2} c \left(-1+\frac{e^{-2 i \beta } (\mu -3)}{\mu -1}\right)
\end{eqnarray}

We would like this coefficient to be zero so that $f(\eta)$ decreases at infinity.

The solution of this complex equation for $\beta, \mu$ is
\begin{eqnarray}
\begin{cases}
\beta =0; &\mu =1 -\frac{c}{a};\\
\beta = \oh \pi; & \mu = 1-\frac{ c}{b};\\
\end{cases}
\end{eqnarray}
From the inequalities between the three eigenvalues $a<b<c$ adding to zero, we get
\begin{eqnarray}
 &&-2 c < a < -\frac{c}{2};\\
 && -\frac{c}{2} < b < c;
\end{eqnarray}
which makes this index $\mu$ limited to the interval
\begin{eqnarray}
\begin{cases}
\beta =0; &\frac{3}{2} < \mu < 3;\\
\beta =\oh \pi; &-\infty < \mu < \infty;
\end{cases}
\end{eqnarray}

We choose the first solution, as in this case, $\mu$ varies in the positive region where there are no singularities.

Now, we observe that at $\beta =0,\oh\pi$, which means $q =0$ there is an extra symmetry in the equation \eqref{formEq}
\begin{eqnarray}
&&C \Ra C^*
\end{eqnarray}

This means that now all four branches of the hyperbola
\begin{eqnarray}
|y| |x|^\mu = \mbox{const};
\end{eqnarray}
are the parts of a single periodic solution $C(\theta)$ (Fig \ref{fig::HyperLoop}).

These four branches define the domain inside the loop $C$  where the holomorphic velocity
\begin{eqnarray}
F_-(\eta) = -2 a \eta;
\end{eqnarray} 

As for the function $F_+$ it is just a negative power
\begin{eqnarray}
&&F_+\left(x \pm\i |x|^{-\mu}\right) = + 2\i c |x|^{-\mu};
\end{eqnarray}
The loop $C(\theta)$ defined by these four branches is a periodic function, with singularities  at $ \theta = k \pi/2$.

Topologically, on a Riemann sphere, these four branches divide the sphere into five regions.
There is an inside region, with the boundary touching Riemann infinity four times. 

\begin{figure}%
    \centering
      \includegraphics[width=\textwidth]{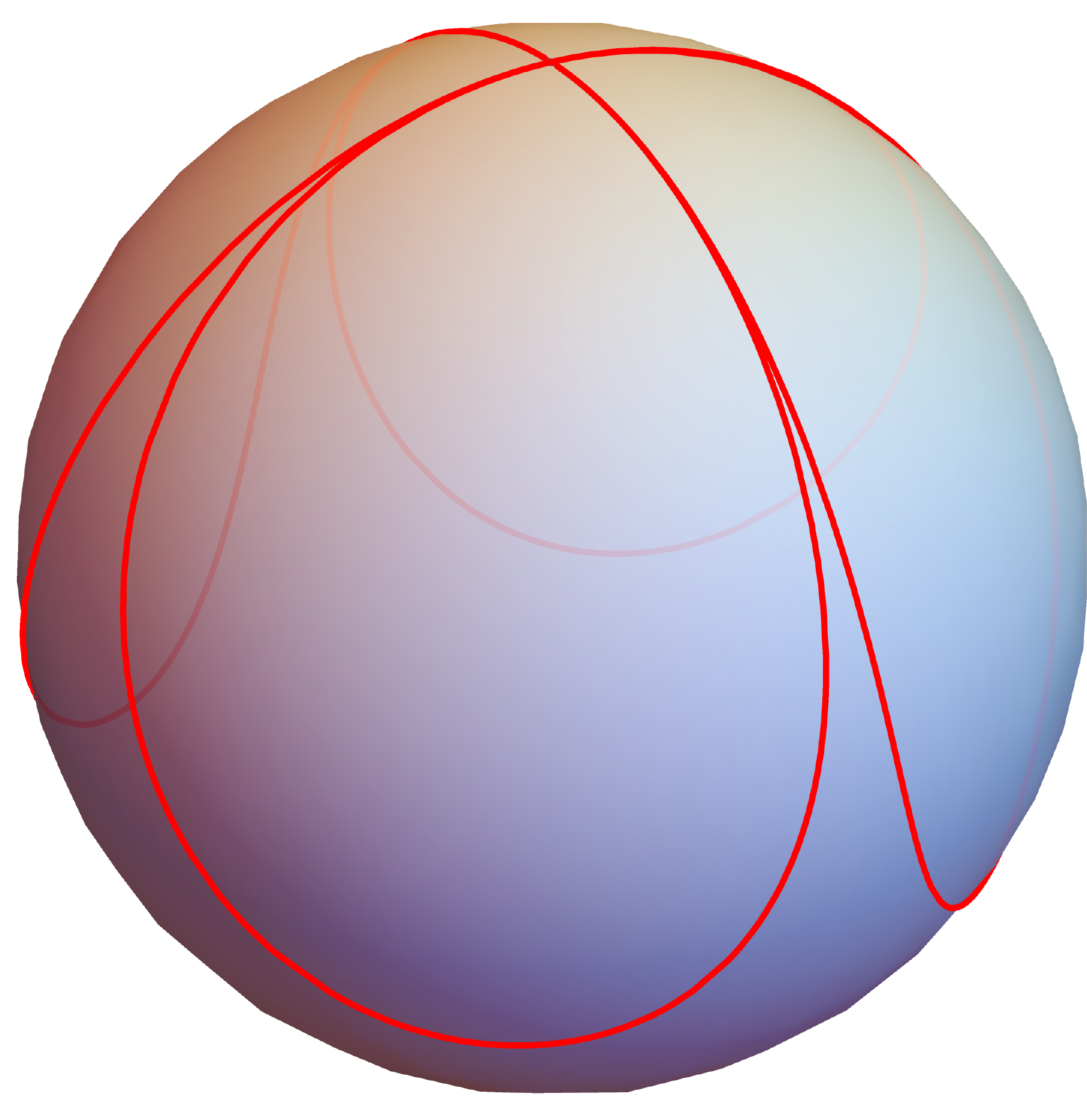}
    \caption{The stereographic projection of four branches of hyperbola on the Riemann sphere.}
    \label{fig::RiemannSphere}
\end{figure}

Each of the four remaining outside regions is bounded by hyperbola,  starting and ending at infinity.
(Fig. \ref{fig::RiemannSphere}).

Let us consider the lower right hyperbola with $y = -x^{-\mu}, x >0$

The values of $F_+(\eta)$ in three other external regions  will be obtained from this one by reflection  against $x$ or $y$ axis.
\begin{eqnarray}
&&F_+(-\eta) = - F_+(\eta);\\
&&F_+(\eta^*) = F_+^*(\eta);
\end{eqnarray}

We can continue the above parametric equation for the function $f$ to the complex plane for the variable $w$, with the cut from $ -\infty $ to $0$. The phase of the multivalued function $w^{-\mu}$ is set to $\arg w^{-\mu} =0, w >0$.
\begin{eqnarray}
&&\eta = w - \i w^{-\mu};\\
&& F_+=  2\i c w^{-\mu};
\end{eqnarray}

This parametric function has a square root singularity at the point where derivative $\d_w \eta = 1 + \i \mu w^{-\mu-1}$ vanishes:
\begin{eqnarray}
w_c = (-\i \mu)^{\frac{1}{\mu+1}}
\end{eqnarray}

For positive $\mu$, this singularity is located in the lower semiplane of $w$, leaving the upper semiplane as a physical domain.

\begin{figure}%
    \centering
      \includegraphics[width=\textwidth]{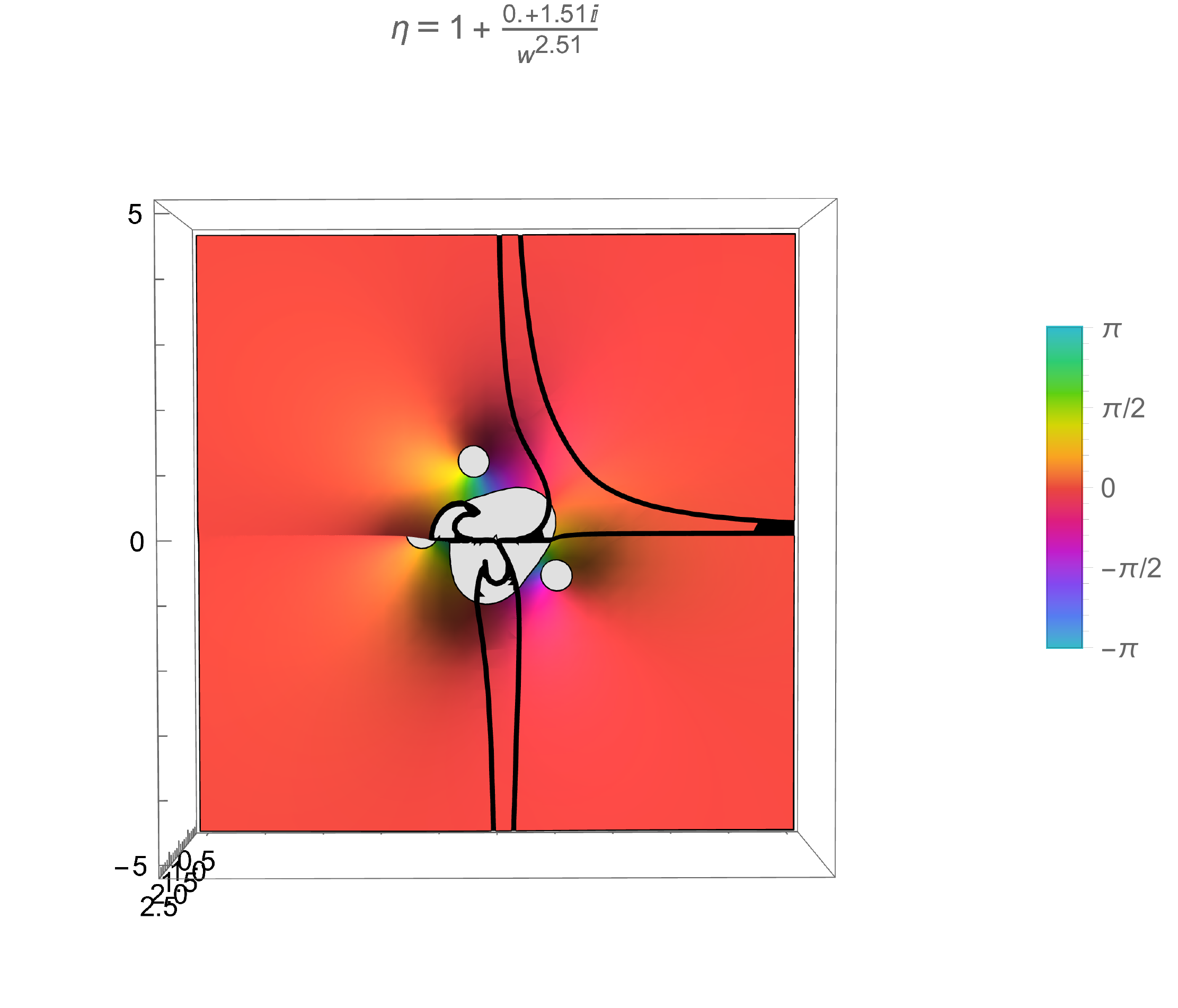}
    \caption{The complex Map3D of the holomorphic function $\eta'(w)= 1 + \i \mu w^{\mu-1}$ for $\mu = 1.51$. The height is $|\eta'|$, the color is $\arg \eta'$. The square root singularities of inverse function $\eta(w)$ is located at the points where $\eta'(w)=0$. They are indicates as a holes on the surface (white circles). The black lines are  described by $(\Re\eta(w))^\mu \Im \eta(w) =\pm1$. The physical region is outside the black line in the first quadrant, and there are no singularities there.}
    \label{fig::DerMap1}
\end{figure}

\begin{figure}%
    \centering
      \includegraphics[width=\textwidth]{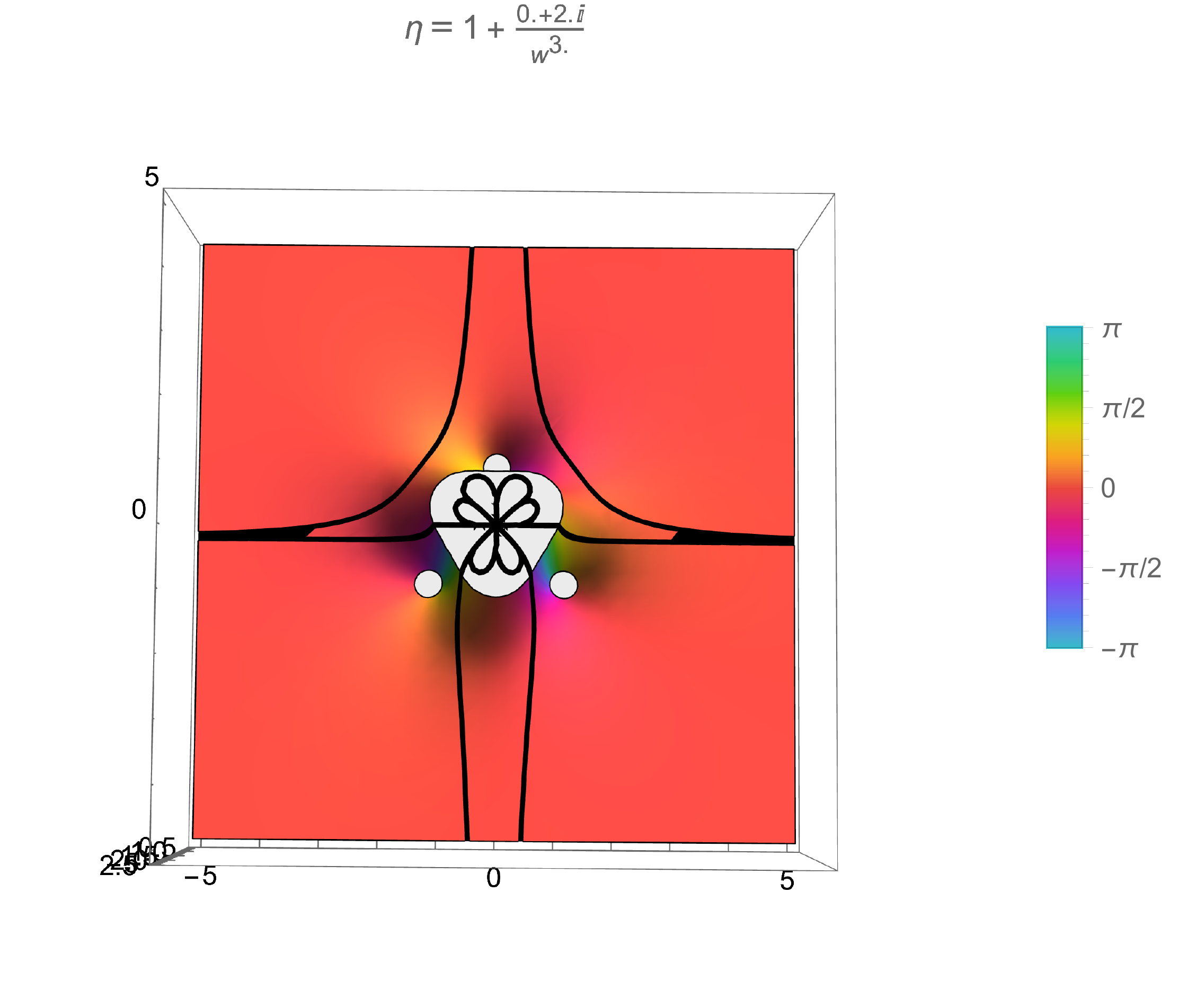}
    \caption{The complex Map3D of the holomorphic function $\eta'(w)= 1 + \i \mu w^{-\mu-1}$ for $\mu = 2$. The height is $|\eta'|$, the color is $\arg \eta'$. The square root singularities of inverse function $\eta(w)$ is located at the points where $\eta'(w)=0$. They are indicates as a holes on the surface (white circles). The black lines are  described by $(\Re\eta(w))^\mu \Im \eta(w) =\pm1$. The physical region is outside the black line in the first quadrant, and there are no singularities there.}
    \label{fig::DerMap2}
\end{figure}

\begin{figure}%
    \centering
      \includegraphics[width=\textwidth]{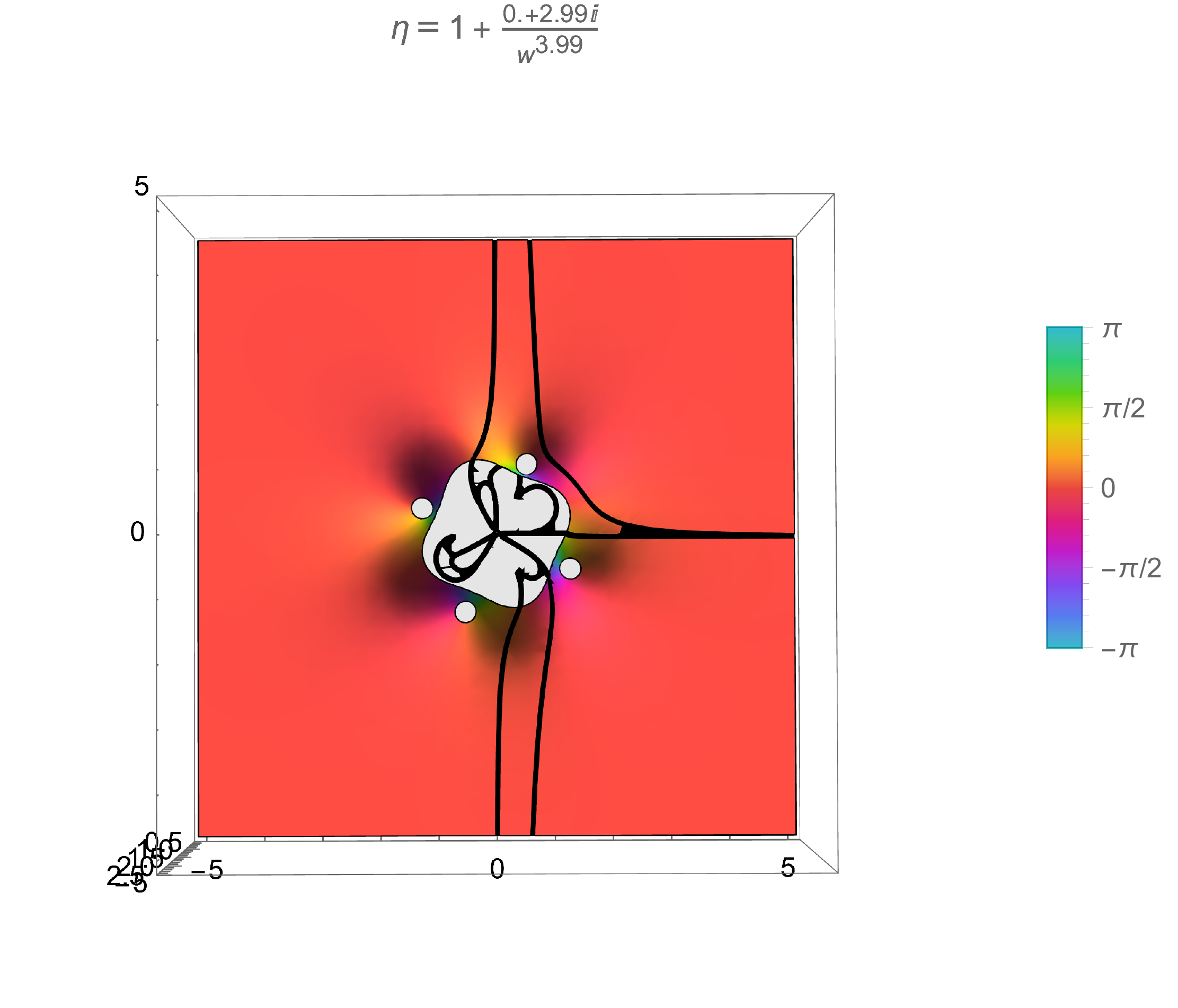}
    \caption{The complex Map3D of the holomorphic function $\eta'(w)= 1 + \i \mu w^{-\mu-1}$ for $\mu = 2.99$. The height is $|\eta'|$, the color is $\arg \eta'$. The square root singularities of inverse function $\eta(w)$ is located at the points where $\eta'(w)=0$. They are indicates as a holes on the surface (white circles). The black lines are  described by $(\Re\eta(w))^\mu \Im \eta(w) =\pm1$. The physical region is outside the black line in the first quadrant, and there are no singularities there.}
    \label{fig::DerMap3}
\end{figure}

We have drawn the complex maps of the derivative $\d_w \eta = 1 + \i \mu w^{-\mu-1}$ for  $\mu=1.51,2,2.99$ (Fig. \ref{fig::DerMap1},\ref{fig::DerMap2},\ref{fig::DerMap3}).
We indicate the zeros of $\d_w \eta$  as holes on the surface (white circles). 

To check whether these singularities penetrate the physical region, we have drawn in black the boundaries of the physical regions
\begin{eqnarray}
\Re\eta^\mu \Im \eta =\pm1
\end{eqnarray}

We observe that these square root singularities lie outside the physical region. This physical region is above the black line in the first quadrant, and there are no singular points there.

Inside the tube, the velocity is linear, and an extra linear term $ F_-(\eta) = -2 a \eta;$ is potential, so that the incompressibility is preserved
\begin{eqnarray}
V^- = a x - \i b y -2 a (x + \i y) = -a x + \i (c-a) y
\end{eqnarray}

\begin{figure}%
    \centering
      \includegraphics[width=\textwidth]{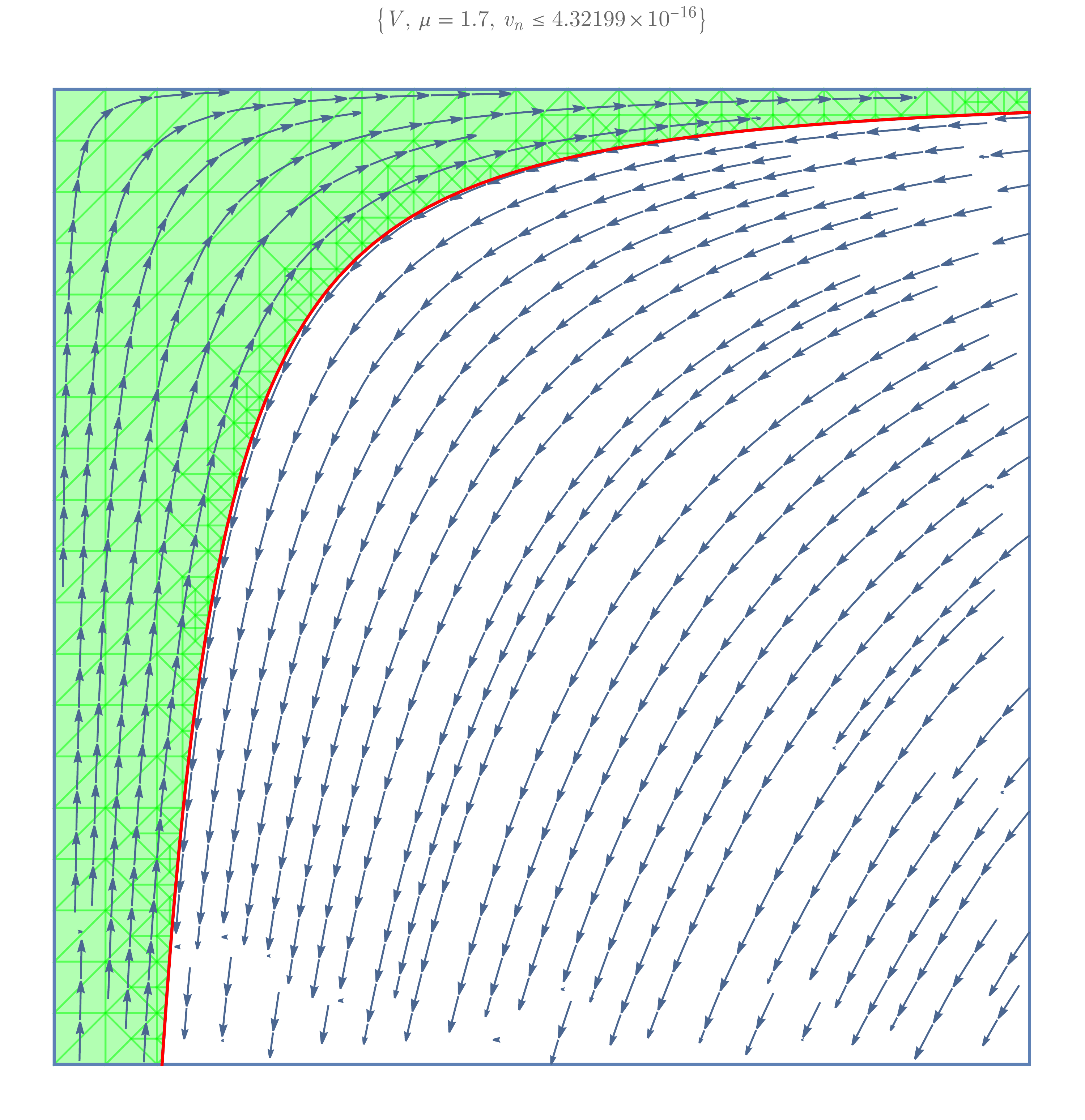}
    \caption{The stream  plot in $IV$ quadrant of $x y $ plane for $\mu = 1.7$. The green fluid is inside, the clear fluid is outside the vortex surface (red).
    The flow in other quadrants can be obtained by reflection against the $x$ and $y$ axes.
    The normal velocity vanishes at the vortex sheet on both sides, with accuracy $\sim 10^{-16}$.}
    \label{fig::Flow}
\end{figure}

\begin{figure}%
    \centering
      \includegraphics[width=\textwidth]{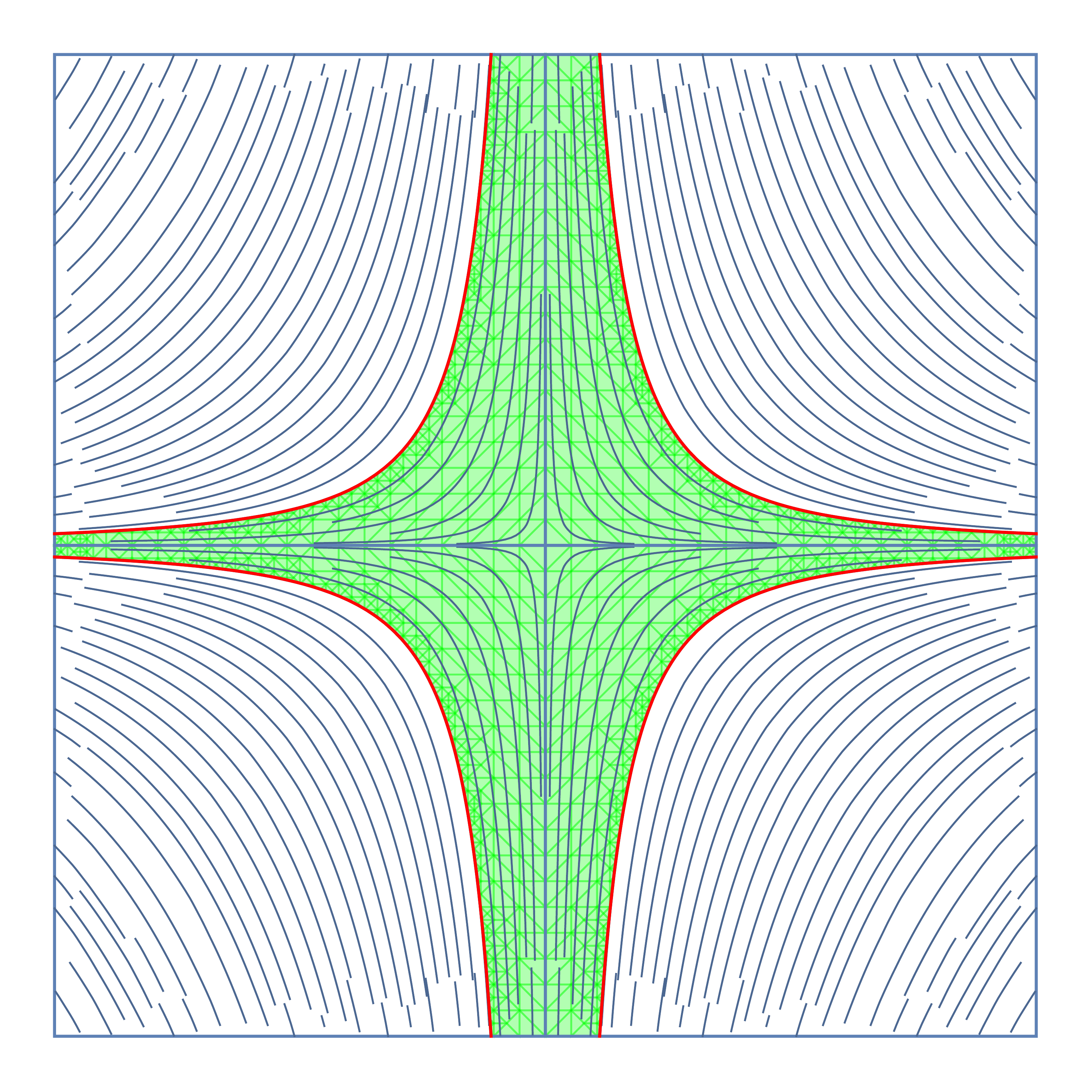}
    \caption{The stream  plot in  $x y $ plane for $\mu = 1.7$. The green fluid is inside, the clear fluid is outside the vortex surface (red)}.
    \label{fig::FlowAllQuads}
\end{figure}

To summarize, the velocity field $\vec v(x,y,z)$ and complex coordinates $\eta = x + \i y$ are parametrized as a function of a complex variable $w$ as follows
\begin{subequations}\label{velocityField}
\begin{eqnarray}
&&v_z = c z;\\
&& v_x - \i v_y = a x - \i b y +  F_\pm\left(\frac{x + \i y}{R_0}\right);\\
&& F_-(\eta) = - 2 a \eta;\\
&& F_+(\eta)=  2\i c R_0 w(\eta)^{-\mu};\; \Re \eta >0, \Im \eta < 0;\\
&& F_+(-\eta) = -F_+(\eta); F_+(\eta^*) = F_+^*(\eta);\\
&&w(\eta): \eta = w - \i w^{-\mu};\\
&&C:  |x|^\mu |y| = R_0^{1 + \mu};\\
&& \mu = 1 - \frac{c}{a};\; \frac{3}{2} < \mu < 3;\\
&& \Delta \Gamma = \oint_C \vec v(\vec r) d \vec r  =0;
\end{eqnarray}
\end{subequations}
We show in Fig. \ref{fig::Flow}, \ref{fig::FlowAllQuads} the streamline plot of this flow in the $x y $ plane.
We performed analytical and numerical computations in \cite{MB15}.

While computing the complex velocity and coordinates using parametric equations, we restricted $w$ to the physical region and populated this region by an irregular grid in angular variables in the $w$ plane. 

After that we used \textit{ListStreamPlot} method of \Mathematica.

The grid resolution does not allow tracking the boundary's immediate vicinity, but we computed the normal velocity numerically at the boundary. It turned out less than $10^{-15}$ on both sides of the sheet.

\section{Viscous cutoff}

\begin{figure}%
    \centering
      \includegraphics[width=\textwidth]{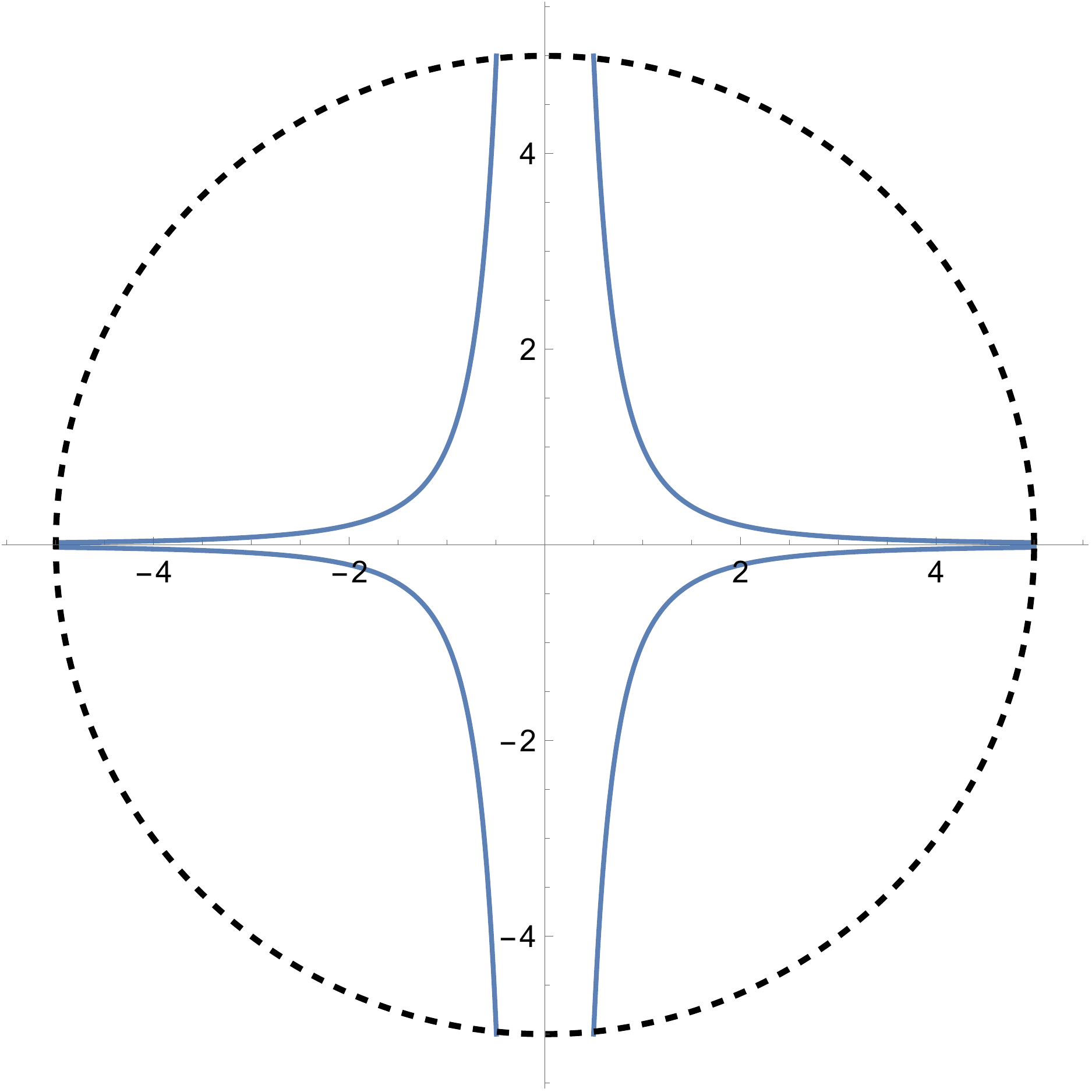}
    \caption{The loop made of four branches of hyperbola.}
    \label{fig::HyperLoop}
\end{figure}

The curve has an infinite length, and it encircles an infinite area because of the infinite cusps at $x =\pm 0, y = \pm 0$
(see Fig.\ref{fig::HyperLoop}).

At a large upper limit $x_{max} \ra \infty$ and small lower limit $x_{min}\ra 0$ the perimeter of the loop goes as
\begin{eqnarray}
&&P =  4 R_0 \int_{x_{min}}^{x_{max}} d x \sqrt{1 + \mu^2 x^{-2(\mu+1)}} \ra 4\left(x_{min}^{-\mu} + x_{max}\right) R_0;
\end{eqnarray}

In the viscous fluid, with finite $\nu$ these infinities will not occur, of course. Our solution applies only as long as the spacing between the two branches of the hyperbola at Fig.\ref{fig::HyperLoop} is much larger than the viscous thickness of the vortex sheet.
\begin{eqnarray}\label{PR0}
&& h = \sqrt{\frac{\nu}{c}};\\
&& x_{min}  = \frac{h}{R_0};\\
&& x_{max} = \left(\frac{R_0}{h}\right)^{\frac{1}{\mu}};\\
&& P \ra  4 R_0\left(\frac{R_0}{h}\right)^\mu + 4R_0\left(\frac{R_0}{h}\right)^{\frac{1}{\mu}}  \ra 4 R_0^{\mu+1}h^{-\mu}
\end{eqnarray}

To keep the perimeter fixed in the extreme turbulent limit, we have to tend the parameter $R_0$ to zero as
\begin{eqnarray}\label{R0Scaling}
&&R_0 = \left(\frac{P}{4}\right)^{\frac{1}{\mu+1}} \left(\frac{\nu}{c}\right)^{\frac{\mu}{2(\mu+1)}} \ra 0;
\end{eqnarray}

The cross section area of the tube
\begin{eqnarray}\label{Area}
\mbox{Area} = 4 R_0^2\int_{x_{min}}^\infty d x x^{-\mu} = \frac{4}{\mu-1} R_0^2 x_{min}^{1-\mu} \propto P h \sim P \left(\frac{\nu}{c}\right)^{\oh}
\end{eqnarray}
\section{Velocity Gap and Circulation}
Once the equation is solved, the parametric solution for $\Gamma$ is straightforward (with factors of $R_0$ restored from dimensional counting)
\begin{eqnarray}\label{eqGamma}
 &&\Gamma = \int (-2 a X X'+2 (a-c) Y Y') d x =  -a R_0^2\left(x^{2}-\mu |x|^{-2 \mu }\right);\\
 &&\Delta V = \frac{\Gamma'}{C'} = -2 a R_0 x\left(1 + \i \mu |x|^{-\mu-1 }\right)
\end{eqnarray}

We have not expected to find such a singular vortex tube, but it satisfies all requirements and must be accepted.

In \cite{M21c}, we appealed to the Brouwer theorem \cite{BrT} to advocate the existence of solutions of the \CVS{} equations. This theorem does not tell us how many fixed points are on a sphere made of the normalized Fourier coefficients in the limit their number going to infinity. Perhaps, there are also some nonsingular solutions.

This singular solution is not normalizable, and it has a vanishing circulation 
\begin{eqnarray}
\Delta \Gamma =  \frac{c R_0^2 }{\mu -1}\left(x^{2}-\mu |x|^{-2 \mu }\right)_{x = -\infty}^{x = \infty} = 0
\end{eqnarray}

Thus, the Brouwer theorem does not apply here -- this is a more general case than the one assumed in \cite{M21c}. On the other hand, we do not need an existence theorem anymore once we have found an analytic solution.

Note that the net circulation would be infinite unless we combine all four branches of our hyperbola into a single closed-loop (with cusps at the real and imaginary axes, but still closed).

\section{Minimizing Euler Energy}

Let us now compute the energy of the vortex surface as a Hamiltonian system \cite{M88,AM89}. 
There is a regular part related to the background strain. This part is not involved in the minimization we are interested in; it depends on $a,b,c$, which are external parameters for our problem.

The internal part of the Hamiltonian is directly related to the potential gap $\Gamma$ we have computed in the previous section.

\begin{eqnarray}
H_{int} = \int_{\vec r_1,\vec r_2 \in \mathcal S}  d \Gamma(\vec r_1) \wedge d \vec r_1  \cdot d \Gamma(\vec r_1) \wedge d \vec r_1 \frac{1}{8 \pi |\vec r_1 - \vec r_2| }
\end{eqnarray}
In our case of cylindrical surface
\begin{eqnarray}
d \Gamma(\vec r_1) \wedge d \vec r_1  = d \Gamma(\theta) \left\{0,0,d z \right\}
\end{eqnarray}
and we have a separation of variables $\theta, z$. 

The integration over $ z_1, z_2 $ provides the total length $L \ra \infty$ of the cylinder times logarithmically divergent integral over $z_1-z_2$.  We limit this integral to the interval  $(-L, L)$ and compute it exactly 
\begin{eqnarray}
\int_{-L}^L \frac{1}{8 \pi  \sqrt{\eta ^2+z^2}} \, dz = \frac{\log \left(\frac{2 L \left(\sqrt{\eta ^2+L^2}+L\right)}{\eta ^2}+1\right)}{8 \pi }
\end{eqnarray}

Then we expand it for large $L$
\begin{eqnarray}
\int_{-L}^L \frac{1}{8 \pi  \sqrt{\eta ^2+z^2}} \, dz  \ra \frac{\log( 2 L) -\log |\eta|}{4 \pi} + O\left(\frac{\eta^2}{L^2}\right)
\end{eqnarray}

Thus we get in our case , with  $\Gamma'(w)$ from \eqref{eqGamma}
\begin{eqnarray}
&&\frac{H_{int}}{L} = \log (2 L) \frac{(\Delta \Gamma)^2 }{4 \pi}-\frac{1}{8 \pi} \Pint_{-\infty}^{\infty} d w_1\Gamma'(w_1) \Pint_{-\infty}^{\infty} d w_2\Gamma'(w_2) \log \abs{\xi_1 -\xi_2};\\
&& \xi_{1,2} = w_{1,2} - \i |w_{1,2}|^{-\mu};\\
&& \Gamma'(w) = \frac{2 c R_0^2 w }{\mu -1}\left(1 +  \mu^2 |w|^{-2 \mu -2}\right)
\end{eqnarray}

At $L \ra \infty$ the first term is the leading one. Minimization of the energy leads to the condition
\begin{eqnarray}
\Delta \Gamma =0
\end{eqnarray}

Therefore, our solution with zero circulation is singled out among other combinations of hyperbolic vortex sheets by the requirement of minimization of energy.

Once the divergent term vanishes, one can compute the rest of the energy. 

The remaining principal value integral over $x_1, x_2$ converges at infinity. In the local limit, when $R_0 \ra 0$, it goes to zero.

We do not see a point in this computation compared to observable quantities such as the energy dissipation and the Wilson loop.

\section{The induced background strain}

What is the physical origin of the constant background strain $W_{\alpha\beta}$ which we used in our solution?

Traditionally, the ad hoc Gaussian random forces are added to the Navier-Stokes equation to simulate the effects of the unknown inner randomness.

We do not think that this beautiful equation needs any crutches; it can walk all by itself. 

In our theory, the random forces come from many remote vortex structures, contributing to the background velocity field via the Biot-Savart law.

These forces are not arbitrary; they are rather self-consistent, like a mean-field in ordinary statistical mechanics.

Let us assume that the space is occupied by some localized \CVS{} structures far from each other.  In other words, let us consider an ideal gas of vortex bubbles. 

We shall see below that the mean size $R_0$ of the surface is small compared to the mean distance $\bar R$ between them in the turbulent limit $\nu \ra 0, \mathcal E = \mbox {const} $. This vanishing size vindicates the assumptions of the low-density ideal gas.

In such an ideal gas, we can neglect the collision of these extended particles, but not the long-range effect of the strain they impose on each other.

The Biot-Savart formula for the velocity field induced by the set of remote localized vorticity bubbles $B$
\begin{eqnarray}
    \vec v(\vec r) = \sum_B\int_B d^3 r' \frac{\vec \omega(\vec r') \times (\vec r' - \vec r)}{4 \pi | \vec r - \vec r'|^3}
\end{eqnarray}
falls off as $1/r^2$ for each bubble, like an electric field from the charged body.

Note that all vortex structures in our infinite volume contribute to this background velocity field, adding up to a large number of small terms at every point in space.

While the \NS{} equation is nonlinear, this relation between the local strain and contributions from each vortex tube is \textbf{exactly linear}, as it follows from the linear Poisson equation relating velocity to vorticity.

Therefore, the interaction between bubbles decreases with distance by the power law, which justifies the ideal gas picture in the case of sparsely distributed vortex tubes.

\begin{center}
    \begin{tikzpicture}
\draw (0, 0) circle (3);
\draw[->, red] (3.0, 0.0) -- (3.18989763692, 0.206882004178);
\draw[->, red] (2.99817248106, 0.104698490108) -- (2.86187640313, 0.444418012577);
\draw[->, red] (2.99269215078, 0.209269421232) -- (2.78392904122, 0.0658099999994);
\draw[->, red] (2.9835656861, 0.313585389803) -- (3.02070562755, 0.837037261956);
\draw[->, red] (2.97080420622, 0.41751930288) -- (3.06995364287, 1.18536928931);
\draw[->, red] (2.95442325904, 0.520944533001) -- (3.0879480934, 0.493598974164);
\draw[->, red] (2.9344428022, 0.623735072453) -- (2.83891271278, 1.06033398555);
\draw[->, red] (2.91088717883, 0.725765686799) -- (3.10473557841, 0.510627909767);
\draw[->, red] (2.88378508781, 0.826912067451) -- (2.5060573585, 0.754345348178);
\draw[->, red] (2.85316954889, 0.927050983125) -- (2.78075252939, 0.901143347442);
\draw[->, red] (2.81907786236, 1.02606042998) -- (2.37399730859, 0.781558686335);
\draw[->, red] (2.7815515637, 1.12381978025) -- (3.09508229995, 1.19150164342);
\draw[->, red] (2.74063637293, 1.22020992923) -- (2.81299485895, 1.3564352197);
\draw[->, red] (2.6963821389, 1.31511344037) -- (2.49941942261, 1.26615212382);
\draw[->, red] (2.64884277858, 1.40841468836) -- (3.36903230154, 1.73842863018);
\draw[->, red] (2.59807621135, 1.5) -- (2.50870498738, 1.17045855207);
\draw[->, red] (2.54414428847, 1.5897577927) -- (2.71827641138, 1.58464664852);
\draw[->, red] (2.48711271767, 1.67757871041) -- (2.18813718767, 1.54686708672);
\draw[->, red] (2.42705098312, 1.76335575688) -- (2.09405794429, 1.66037100993);
\draw[->, red] (2.36403226082, 1.84698442598) -- (1.93252639477, 2.21807804904);
\draw[->, red] (2.29813332936, 1.92836282906) -- (2.43423272507, 1.51347700514);
\draw[->, red] (2.22943447643, 2.00739181908) -- (2.06854921552, 1.97359095614);
\draw[->, red] (2.15801940102, 2.08397511138) -- (1.9258335215, 1.7263424932);
\draw[->, red] (2.08397511138, 2.15801940102) -- (1.75565565496, 1.91907655062);
\draw[->, red] (2.00739181908, 2.22943447643) -- (1.66531344206, 1.99636529281);
\draw[->, red] (1.92836282906, 2.29813332936) -- (1.96566980134, 1.94379809258);
\draw[->, red] (1.84698442598, 2.36403226082) -- (1.70487875929, 2.37829869);
\draw[->, red] (1.76335575688, 2.42705098312) -- (1.62182528237, 2.21848916138);
\draw[->, red] (1.67757871041, 2.48711271767) -- (1.70293332919, 2.85993204194);
\draw[->, red] (1.5897577927, 2.54414428847) -- (1.63460823423, 2.25213441541);
\draw[->, red] (1.5, 2.59807621135) -- (1.51237421172, 2.36254630876);
\draw[->, red] (1.40841468836, 2.64884277858) -- (1.56044917181, 2.62350283287);
\draw[->, red] (1.31511344037, 2.6963821389) -- (1.4326193314, 2.58649487031);
\draw[->, red] (1.22020992923, 2.74063637293) -- (1.29146381578, 2.63264373772);
\draw[->, red] (1.12381978025, 2.7815515637) -- (0.613253182305, 2.8114841836);
\draw[->, red] (1.02606042998, 2.81907786236) -- (1.03497921441, 2.67324867851);
\draw[->, red] (0.927050983125, 2.85316954889) -- (0.551482030987, 3.2386645256);
\draw[->, red] (0.826912067451, 2.88378508781) -- (0.580443305142, 3.09117113964);
\draw[->, red] (0.725765686799, 2.91088717883) -- (0.863825020463, 2.69147350839);
\draw[->, red] (0.623735072453, 2.9344428022) -- (0.709160938294, 2.568562005);
\draw[->, red] (0.520944533001, 2.95442325904) -- (0.185063664054, 3.15727266888);
\draw[->, red] (0.41751930288, 2.97080420622) -- (0.477598819342, 3.01834760387);
\draw[->, red] (0.313585389803, 2.9835656861) -- (0.0527268101921, 2.69929557543);
\draw[->, red] (0.209269421232, 2.99269215078) -- (-0.145300820942, 3.36346044384);
\draw[->, red] (0.104698490108, 2.99817248106) -- (0.304621265325, 2.71379047122);
\draw[->, red] (1.83697019872e-16, 3.0) -- (-0.0732179374315, 3.62894708093);
\draw[->, red] (-0.104698490108, 2.99817248106) -- (-0.298047562323, 2.85671684955);
\draw[->, red] (-0.209269421232, 2.99269215078) -- (-0.24374159757, 3.00338661766);
\draw[->, red] (-0.313585389803, 2.9835656861) -- (-0.0184572014415, 3.41722939486);
\draw[->, red] (-0.41751930288, 2.97080420622) -- (-0.352661474172, 2.73506271443);
\draw[->, red] (-0.520944533001, 2.95442325904) -- (-1.67682304378, 2.60410098578);
\draw[->, red] (-0.623735072453, 2.9344428022) -- (-0.226344352811, 2.7172102314);
\draw[->, red] (-0.725765686799, 2.91088717883) -- (-0.874320159866, 2.67885694643);
\draw[->, red] (-0.826912067451, 2.88378508781) -- (-0.930387802353, 2.53269518768);
\draw[->, red] (-0.927050983125, 2.85316954889) -- (-1.13239640231, 2.79620127612);
\draw[->, red] (-1.02606042998, 2.81907786236) -- (-0.948619372619, 2.61302134644);
\draw[->, red] (-1.12381978025, 2.7815515637) -- (-1.21845112514, 2.9238346055);
\draw[->, red] (-1.22020992923, 2.74063637293) -- (-1.21402359492, 2.86791433052);
\draw[->, red] (-1.31511344037, 2.6963821389) -- (-1.59018092296, 2.71650729288);
\draw[->, red] (-1.40841468836, 2.64884277858) -- (-1.40797553653, 2.65572636607);
\draw[->, red] (-1.5, 2.59807621135) -- (-1.23848197872, 2.63974165954);
\draw[->, red] (-1.5897577927, 2.54414428847) -- (-1.57343114712, 2.91323923811);
\draw[->, red] (-1.67757871041, 2.48711271767) -- (-1.51732473682, 2.34852873138);
\draw[->, red] (-1.76335575688, 2.42705098312) -- (-2.06223149649, 2.76621785129);
\draw[->, red] (-1.84698442598, 2.36403226082) -- (-1.96648420569, 2.72176017255);
\draw[->, red] (-1.92836282906, 2.29813332936) -- (-1.68116668044, 2.38792019914);
\draw[->, red] (-2.00739181908, 2.22943447643) -- (-1.98427700146, 2.207824065);
\draw[->, red] (-2.08397511138, 2.15801940102) -- (-2.27486540816, 2.23729426669);
\draw[->, red] (-2.15801940102, 2.08397511138) -- (-2.58464805536, 1.67631550285);
\draw[->, red] (-2.22943447643, 2.00739181908) -- (-2.14390493528, 2.12149980891);
\draw[->, red] (-2.29813332936, 1.92836282906) -- (-2.62011809966, 1.92867779322);
\draw[->, red] (-2.36403226082, 1.84698442598) -- (-2.03101397202, 1.4033297666);
\draw[->, red] (-2.42705098312, 1.76335575688) -- (-2.49172131703, 1.64731954638);
\draw[->, red] (-2.48711271767, 1.67757871041) -- (-2.68711199588, 1.21852721685);
\draw[->, red] (-2.54414428847, 1.5897577927) -- (-2.86847556942, 1.34749171723);
\draw[->, red] (-2.59807621135, 1.5) -- (-2.69346300154, 1.4660419061);
\draw[->, red] (-2.64884277858, 1.40841468836) -- (-2.67130130732, 1.38862466366);
\draw[->, red] (-2.6963821389, 1.31511344037) -- (-3.16709170952, 1.45366002775);
\draw[->, red] (-2.74063637293, 1.22020992923) -- (-2.31560474605, 1.14453440601);
\draw[->, red] (-2.7815515637, 1.12381978025) -- (-2.94191152978, 1.43498732059);
\draw[->, red] (-2.81907786236, 1.02606042998) -- (-2.74434057758, 1.07006003882);
\draw[->, red] (-2.85316954889, 0.927050983125) -- (-2.69263106898, 1.08978634707);
\draw[->, red] (-2.88378508781, 0.826912067451) -- (-2.72734777215, 0.478442701806);
\draw[->, red] (-2.91088717883, 0.725765686799) -- (-2.84168325018, 0.830974929636);
\draw[->, red] (-2.9344428022, 0.623735072453) -- (-2.83489568062, 0.444299979843);
\draw[->, red] (-2.95442325904, 0.520944533001) -- (-2.92014840509, 0.428199685013);
\draw[->, red] (-2.97080420622, 0.41751930288) -- (-2.66383622494, 0.406924869236);
\draw[->, red] (-2.9835656861, 0.313585389803) -- (-2.79707618979, 0.109554742508);
\draw[->, red] (-2.99269215078, 0.209269421232) -- (-3.03547972658, 0.037666203143);
\draw[->, red] (-2.99817248106, 0.104698490108) -- (-2.64949550051, -0.0347696612005);
\draw[->, red] (-3.0, 3.67394039744e-16) -- (-3.04347529899, 0.0247195154345);
\draw[->, red] (-2.99817248106, -0.104698490108) -- (-3.14802135357, 0.136428384273);
\draw[->, red] (-2.99269215078, -0.209269421232) -- (-2.8633154347, -0.114792426255);
\draw[->, red] (-2.9835656861, -0.313585389803) -- (-2.85779839474, -0.299630149963);
\draw[->, red] (-2.97080420622, -0.41751930288) -- (-3.13178657115, -0.574847155054);
\draw[->, red] (-2.95442325904, -0.520944533001) -- (-2.91876244475, -0.405922871455);
\draw[->, red] (-2.9344428022, -0.623735072453) -- (-3.34436009073, -0.671144984914);
\draw[->, red] (-2.91088717883, -0.725765686799) -- (-3.22652305699, -0.594522096622);
\draw[->, red] (-2.88378508781, -0.826912067451) -- (-2.88965284355, -0.791280875084);
\draw[->, red] (-2.85316954889, -0.927050983125) -- (-2.57329714466, -1.18768088164);
\draw[->, red] (-2.81907786236, -1.02606042998) -- (-2.31548993148, -0.955560814495);
\draw[->, red] (-2.7815515637, -1.12381978025) -- (-2.64306314606, -1.43357356318);
\draw[->, red] (-2.74063637293, -1.22020992923) -- (-2.79556284294, -1.02477172835);
\draw[->, red] (-2.6963821389, -1.31511344037) -- (-2.86837394899, -0.929545490032);
\draw[->, red] (-2.64884277858, -1.40841468836) -- (-2.2928484287, -1.21347068465);
\draw[->, red] (-2.59807621135, -1.5) -- (-2.9452681636, -2.33491134724);
\draw[->, red] (-2.54414428847, -1.5897577927) -- (-2.38485580346, -1.38392513667);
\draw[->, red] (-2.48711271767, -1.67757871041) -- (-2.42693193622, -1.71005883878);
\draw[->, red] (-2.42705098312, -1.76335575688) -- (-2.58021736489, -2.05362842271);
\draw[->, red] (-2.36403226082, -1.84698442598) -- (-2.57753029198, -2.01808049085);
\draw[->, red] (-2.29813332936, -1.92836282906) -- (-2.05897015018, -1.71848590207);
\draw[->, red] (-2.22943447643, -2.00739181908) -- (-2.44570276606, -2.25970871455);
\draw[->, red] (-2.15801940102, -2.08397511138) -- (-2.40522641721, -2.57239496883);
\draw[->, red] (-2.08397511138, -2.15801940102) -- (-2.22804190616, -2.15804774161);
\draw[->, red] (-2.00739181908, -2.22943447643) -- (-2.26571530382, -2.75787491001);
\draw[->, red] (-1.92836282906, -2.29813332936) -- (-2.05874326567, -2.76354745826);
\draw[->, red] (-1.84698442598, -2.36403226082) -- (-1.87992422083, -2.72079287842);
\draw[->, red] (-1.76335575688, -2.42705098312) -- (-1.9130533174, -2.41974275229);
\draw[->, red] (-1.67757871041, -2.48711271767) -- (-1.99541526503, -2.9918263612);
\draw[->, red] (-1.5897577927, -2.54414428847) -- (-1.54597838458, -2.17331898586);
\draw[->, red] (-1.5, -2.59807621135) -- (-1.53108938192, -2.24666773134);
\draw[->, red] (-1.40841468836, -2.64884277858) -- (-0.853249123831, -2.6063190776);
\draw[->, red] (-1.31511344037, -2.6963821389) -- (-1.18136858564, -2.81789828092);
\draw[->, red] (-1.22020992923, -2.74063637293) -- (-1.43055933258, -2.70353724505);
\draw[->, red] (-1.12381978025, -2.7815515637) -- (-0.798275008208, -2.85745596864);
\draw[->, red] (-1.02606042998, -2.81907786236) -- (-0.984965307961, -2.14911041462);
\draw[->, red] (-0.927050983125, -2.85316954889) -- (-0.857307365463, -2.70898721676);
\draw[->, red] (-0.826912067451, -2.88378508781) -- (-0.836690279356, -3.13049458975);
\draw[->, red] (-0.725765686799, -2.91088717883) -- (-0.641349468948, -2.66763377287);
\draw[->, red] (-0.623735072453, -2.9344428022) -- (-1.1357114996, -2.99716954073);
\draw[->, red] (-0.520944533001, -2.95442325904) -- (-0.803722923223, -3.10453424292);
\draw[->, red] (-0.41751930288, -2.97080420622) -- (-0.386091178005, -2.86062802752);
\draw[->, red] (-0.313585389803, -2.9835656861) -- (-0.28277270597, -3.09864383986);
\draw[->, red] (-0.209269421232, -2.99269215078) -- (-0.516166908905, -2.85363952263);
\draw[->, red] (-0.104698490108, -2.99817248106) -- (0.0210513862726, -2.97429127975);
\draw[->, red] (-5.51091059616e-16, -3.0) -- (0.0242713301868, -3.52815135671);
\draw[->, red] (0.104698490108, -2.99817248106) -- (0.433559917904, -2.87756960072);
\draw[->, red] (0.209269421232, -2.99269215078) -- (0.0832811303765, -2.75274296472);
\draw[->, red] (0.313585389803, -2.9835656861) -- (0.625462350184, -3.47791128908);
\draw[->, red] (0.41751930288, -2.97080420622) -- (-0.215427846601, -2.91674753389);
\draw[->, red] (0.520944533001, -2.95442325904) -- (0.698746135555, -3.03995003002);
\draw[->, red] (0.623735072453, -2.9344428022) -- (0.381825012443, -3.45095631867);
\draw[->, red] (0.725765686799, -2.91088717883) -- (0.753334033268, -3.01887033499);
\draw[->, red] (0.826912067451, -2.88378508781) -- (0.705564136556, -2.92941724919);
\draw[->, red] (0.927050983125, -2.85316954889) -- (1.41113455115, -3.40534200591);
\draw[->, red] (1.02606042998, -2.81907786236) -- (1.07451301987, -3.03563601835);
\draw[->, red] (1.12381978025, -2.7815515637) -- (0.957630503643, -2.71400528226);
\draw[->, red] (1.22020992923, -2.74063637293) -- (1.00278291385, -2.15590056526);
\draw[->, red] (1.31511344037, -2.6963821389) -- (1.11584795112, -2.48156714269);
\draw[->, red] (1.40841468836, -2.64884277858) -- (1.47229470996, -2.42997355565);
\draw[->, red] (1.5, -2.59807621135) -- (1.36543984678, -2.9415074817);
\draw[->, red] (1.5897577927, -2.54414428847) -- (1.92005553603, -2.0135324779);
\draw[->, red] (1.67757871041, -2.48711271767) -- (1.60455737283, -2.49062004187);
\draw[->, red] (1.76335575688, -2.42705098312) -- (1.5606572359, -2.46252841489);
\draw[->, red] (1.84698442598, -2.36403226082) -- (1.44807230237, -2.00682935572);
\draw[->, red] (1.92836282906, -2.29813332936) -- (2.18815057107, -2.21852192352);
\draw[->, red] (2.00739181908, -2.22943447643) -- (2.15059591715, -2.43092969499);
\draw[->, red] (2.08397511138, -2.15801940102) -- (2.37654352353, -1.40822174014);
\draw[->, red] (2.15801940102, -2.08397511138) -- (1.95867211726, -2.51284844011);
\draw[->, red] (2.22943447643, -2.00739181908) -- (1.84163202633, -2.31766478478);
\draw[->, red] (2.29813332936, -1.92836282906) -- (2.85993170839, -1.87707818054);
\draw[->, red] (2.36403226082, -1.84698442598) -- (1.99752151071, -1.72229460185);
\draw[->, red] (2.42705098312, -1.76335575688) -- (1.8965835219, -1.62996390383);
\draw[->, red] (2.48711271767, -1.67757871041) -- (2.80647737019, -1.45189383519);
\draw[->, red] (2.54414428847, -1.5897577927) -- (2.5273239967, -1.66408286705);
\draw[->, red] (2.59807621135, -1.5) -- (2.43947066427, -1.72825089551);
\draw[->, red] (2.64884277858, -1.40841468836) -- (2.42336328609, -1.0644063492);
\draw[->, red] (2.6963821389, -1.31511344037) -- (2.77666951601, -1.35593722929);
\draw[->, red] (2.74063637293, -1.22020992923) -- (2.28728115697, -1.86575140416);
\draw[->, red] (2.7815515637, -1.12381978025) -- (2.85856133106, -0.824183760798);
\draw[->, red] (2.81907786236, -1.02606042998) -- (2.64034459611, -1.40192744459);
\draw[->, red] (2.85316954889, -0.927050983125) -- (2.52983147664, -1.1933025616);
\draw[->, red] (2.88378508781, -0.826912067451) -- (3.3150543001, -0.537321731144);
\draw[->, red] (2.91088717883, -0.725765686799) -- (3.20664407828, -1.07084249221);
\draw[->, red] (2.9344428022, -0.623735072453) -- (2.37786593194, -0.508453913022);
\draw[->, red] (2.95442325904, -0.520944533001) -- (3.36526729818, -0.648952283581);
\draw[->, red] (2.97080420622, -0.41751930288) -- (2.75417204049, -0.460270323914);
\draw[->, red] (2.9835656861, -0.313585389803) -- (3.62780016563, -0.297775276693);
\draw[->, red] (2.99269215078, -0.209269421232) -- (2.82528433849, -0.462939878866);
\draw[->, red] (2.99817248106, -0.104698490108) -- (2.85837812702, -0.319398346917);
\node[inner sep=0pt] (vortex) at (0.5, 0.5) {\includegraphics[width=2.5cm]{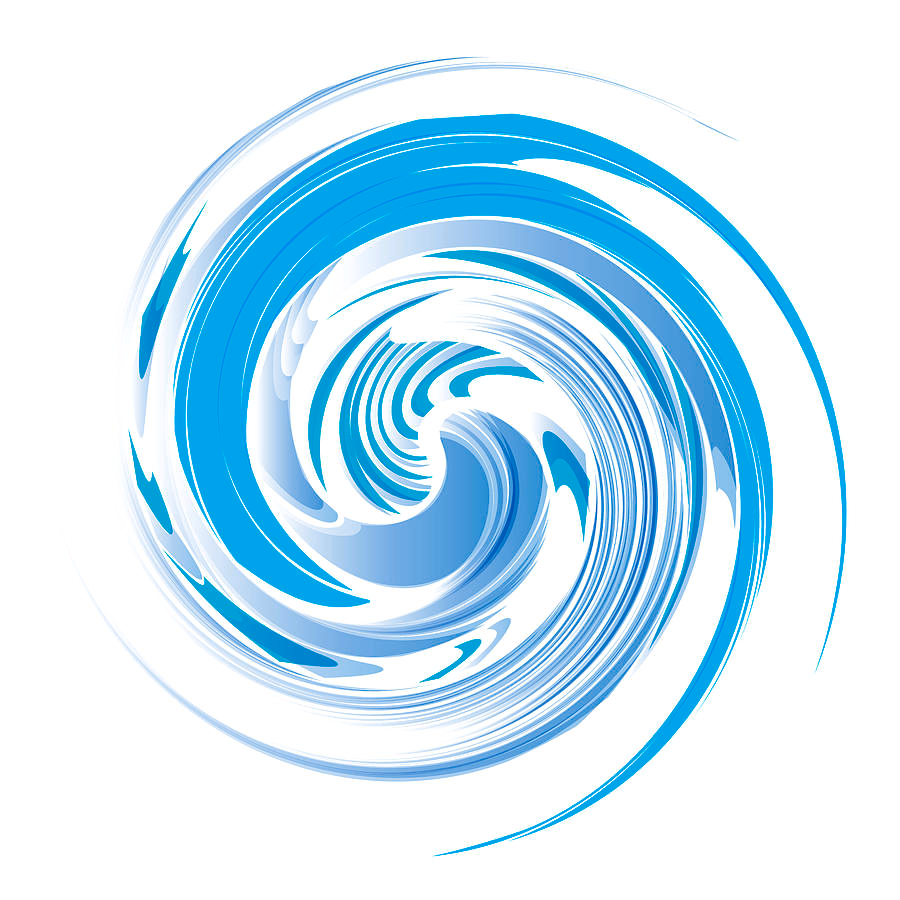}};
\end{tikzpicture}

\end{center}
This picture symbolizes the vortex tube under consideration (blue vortex symbol) surrounded by other remote tubes on a large sphere (orange arrows). The arrows symbolize the directions of these remote tubes, which follow the local strain main axis and point in random directions.

If there are many such bubbles distributed in space with small but finite density, we would have the "night sky paradox."  The bubbles spread on the far away sphere will compensate the inverse distance squared for a divergent distribution like $\int R^2 d R/R^2$.

This estimate is, of course, wrong, as the velocity contributions from various bubbles are uncorrelated, so there is no coherent mean velocity. 

Moreover, a Galilean transformation would remove the finite background velocity, so it does not have any physical effects. 

However, with the strain, there is another story.
Strain coming from remote vortex bubbles
\begin{eqnarray}\label{BSStrain}
    W_{\alpha\beta}(\vec r) = \oh e_{\alpha\mu\gamma}\dbe \dga  \sum_B\int_B d^3 r' \frac{ \omu(\vec r') }{4 \pi | \vec r - \vec r'|} + \left\{ \alpha \leftrightarrow \beta \right\};
\end{eqnarray}
falls off as $1/r^3$, and this time, there could be a mean value $\bar W$, coming from a large number of random terms  from various bubbles with distribution $ R^2 d R/R^3 \sim d R/R$.

The space symmetry arguments and some refined arguments we present in the next section tell us that averaging over the directions of the bubble centers $\vec R = \vec r'-\vec r$ completely cancels this mean value.

The Central Limit Theorem suggests (within our ideal vortex gas model) that such a strain would be a Gaussian tensor variable, satisfying the normal distribution of a symmetric traceless matrix with zero mean (see Appendix B.)
\begin{eqnarray}\label{GaussTensor}
    d P_\sigma(W) \propto \prod_i d W_{i i} \prod_{i < j} d W _{i j} \delta\left(\sum_i W_{i i}\right)\exp{-\frac{\tr W^2}{2 \sigma^2}}
\end{eqnarray}
The parameter $\sigma$ is related to the mean mean square  of the random matrix. In $n$ dimensional space
\begin{eqnarray}\label{TraceRel}
\frac{(n+2)(n-1)}{2} \sigma^2 = \VEV{tr W^2}
\end{eqnarray}
The Gaussian random matrices were studied extensively in physics and mathematics. For example, in \cite{RSM} the distribution of Gaussian random symmetric matrix (Gaussian Orthogonal Ensemble, $GOE(n)$) is presented.

We achieve the extra condition of zero matrix trace by inserting the delta function of the matrix trace into the invariant measure. This projection preserves the measure's $O(n)$ symmetry as the trace is invariant to orthogonal transformations. We could not find any references for this straightforward extension of the $GOE(n)$ to the space of traceless symmetric matrices.

Separating $SO_3$ rotations $\Omega \in S_2$, we have the measure for eigenvalues $a,b,c$:
\begin{eqnarray}\label{Pab}
    &&d P_\sigma(W) =\frac{1}{4 \pi}d \Omega  d a d b d c \delta(a + b + c) P_\sigma(a,b,c);\\
    && P_\sigma(a,b,c)=  \sqrt{\frac{3}{\pi }} \theta(b-a) \theta(c-b) (b-a)(c-a)(c-b)\nonumber\\
    && \exp{-\frac{a^2 + b^2 + c^2}{2\sigma^2} }
\end{eqnarray}

\section{ Energy dissipation and its distribution}
As we noticed in the previous work \cite{M21c} the total surface dissipation is conserved on \CVS{} surfaces.
\begin{eqnarray}\label{Etotal}
\mathcal E_{tot} = \sum_{\mathcal S} \mathcal E_{\mathcal S} = \mbox{const}
\end{eqnarray}

Without CVS as a stability condition, the surface dissipation itself would not be an integral of motion. The energy would leak from the vortex surfaces and dissipate in the rest of the volume. Thus, the CVS condition is a necessary part of the vortex sheet turbulence.

While the total dissipation is conserved, the individual contributions to this sum from each tube are not.  The long-term interactions between the vortex tubes, arising due to the Gaussian fluctuations of the background strain, lead to the statistical distribution of the energy dissipation of an individual tube.

From analogy with the Gibbs-Boltzmann statistical mechanics, one would expect that the dissipation distribution would come out exponential, with some effective temperature. This hypothesis was put forward in our previous work.

However, the interaction between our tubes is different from that of the Gibbs mechanics. While the background strain is a Gaussian (matrix) variable, the shapes of the tubes and the corresponding dissipation are not. 

These tubes in our incompressible fluid instantly adjust to the realization of the random background strain.
This adjustment is described by our exact solutions of the Euler equations with the \CVS{} boundary conditions.

The general formula \cite{M21c} for the surface dissipation reads
\begin{eqnarray}\label{DissipationStrain}
  \mathcal E_{\mathcal S} =  \frac{\sqrt{\nu}}{ 2\sqrt{\pi} } \int_{\mathcal S} d S\sqrt{-\hat S_{n n} } (\Delta \vec v)^2;
\end{eqnarray}
where $\int d S$ refers to the surface integral and $\hat S_{n n}$ is the normal component of the local strain.

For our hyperbolic loop solution  the energy dissipation integral reduces to the following expression (where we restore the implied spatial scale $R_0$ and expressed the cutoffs in terms of the perimeter $P$ of the loop by \eqref{R0Scaling})
\begin{eqnarray}\label{Dissipation}
  && \frac {\mathcal E} { L \sqrt {\nu }} = 4 \frac{2 a^2 \sqrt{c}}{\sqrt{\pi } }  R_0^3\int_{x_{min}}^{x_{max}} \, d w \, w^2 \abs{1- \i \mu  w^{-\mu -1}}^3 \ra  P^3 \frac{ a^2 \sqrt{c}}{24\sqrt{\pi }}
\end{eqnarray}
The normalized distribution $W(\zeta)$ for the scaling variable $\zeta = \frac{a^2 \sqrt{c}}{\sigma^{\frac{5}{2}}}$ takes the form \cite{MB3}
\begin{eqnarray}\label{DisPDF}
&& W(\zeta)=2 \zeta^{9/5} \sqrt{\frac{3}{\pi }}\nonumber\\
&&\int_{2^{-\ofi}}^{2^{\ofi}}d y \frac{\left(2-y^5\right) \left(y^5+1\right) \left(2 y^5-1\right) }{y^{14}} \exp{\frac{\left(-y^{10}+y^5-1\right) \zeta^{4/5}}{y^8}}\\
&&\zeta = \frac { 24 \sqrt{\pi} \mathcal E} { L P^3\sqrt {\nu \sigma^5}};
\end{eqnarray}

The expectation value of this scaling variable equals to
\begin{eqnarray}
\bar \zeta = 4.90394
\end{eqnarray}

\begin{figure}%
    \centering
      \includegraphics[width=\textwidth]{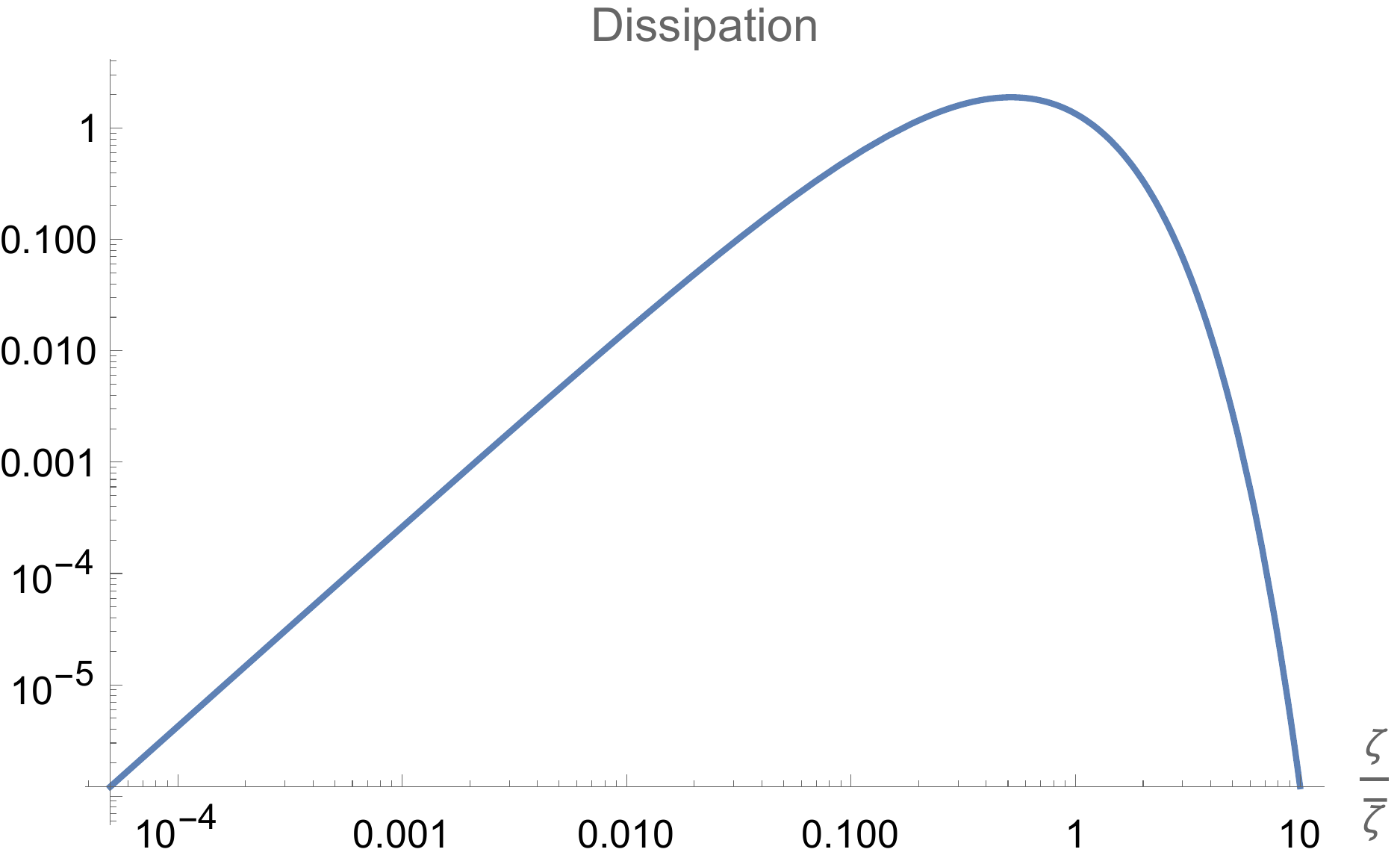}
    \caption{The energy dissipation PDF (fixed perimeter) in log-log scale}
    \label{fig::Dissipation}
\end{figure}

We show at Fig.\ref{fig::Dissipation} the log-log plot of this distribution.

This $W(\zeta)$ is a completely universal function. We would verify this prediction when the distribution of energy dissipation and tube sizes in numerical or real experiments in the extreme turbulent regime will become available.

The perimeter $P$ of the loop remains a free parameter of our theory. We need some extra restrictions to find the distribution of these perimeters. This extra restriction of the fixed perimeter of the cross-section makes it quite tedious to compare our distribution of the energy dissipation with numerical simulation.

This comparison would be a subject of the subsequent numerical project, using supercomputer resources to analyze the database of the turbulent flow from \cite{S19}.

\section{Dilute gas of vortex bubbles in mean-field approximation}

Let us elaborate on this idea of a dilute gas of vortex bubbles and compute the strain variance. 

Consider a large number of independent vortex bubbles,  sparsely distributed in the 3D volume. 

The net strain near this surface will come from the \BS{} formula, which we expand at large distances
\begin{eqnarray}
 &&W_{\alpha\beta}(\vec r) \ra \oh e_{\alpha\mu\gamma} \left(\Omega_{\mu} \dbe \dga +  \Omega_{\mu \lambda }\dbe \dga \d_\lambda+ \dots\right) \frac{1}{4 \pi |\vec r |} + \left\{ \alpha \leftrightarrow \beta \right\};\\
 &&\Omega_{\mu } = \int_B d^3 r' \omu(\vec r');\\
 &&\Omega_{\mu \lambda } = \int_B d^3 r' \omu(\vec r') r'_\lambda 
\end{eqnarray}

The comment is in order. At zero viscosity, the surface is infinite, and the integrals for multipole moments $\Omega_{\mu\dots}$ diverge. 

We consider the case of small but finite viscosity when the surface is finite due to the viscous cutoff at the cusps. 
We assume the distance $r$ to be much larger than the size of this clipped surface. This assumption will be vindicated below.

The contribution to the strain from each remote vortex blob will be linearly related to these multipole moments of vorticity.

These vectors and tensors are random variables with zero mean\footnote{The directions of these vectors in our exact solution coincide with one of the eigenvectors of the local strain tensor at the location of that particular CVS{} surface. Assuming random locations in space, so will be the strain tensors and so will be the vorticity vectors (directed along the symmetry axis of the cylindrical solution).}, in addition to the random locations on a sphere, which is why we expect the Central Limit Theorem to apply here).

The vorticity for each vortex bubble $S$ is given by a surface integral\cite{M88, AM89}
\begin{eqnarray}
  &&\onu(\vec r)  = \int_S d \Gamma \wedge d r'_\nu  \delta^3(\vec r - \vec r');\\
  && \Omega_{\mu  } =  \int_S d \Gamma \wedge d r_\mu  ;\\
  && \Omega_{\mu \lambda } =  \int_S d \Gamma \wedge d r_\mu  r_\lambda 
\end{eqnarray}

We get exactly zero when averaged over directions of the position vector $\vec r $ of the bubble on the large sphere. We verified that up to the fourth term by symbolic integration \cite{MB8}. There is, of course, a general reason for these cancellations.

The rotational average of the multiple derivative matrix has only one totally symmetric symmetric tensor structure
\begin{eqnarray}
  &&T_{\mu_1,\dots\mu_n} = \VEV{\d_{\mu_1}, \dots \d_{\mu_n} \frac{1}{|\vec r|}}_{\vec r \in S_2} = \nonumber\\
  && = C \left( \delta_{\mu_1\mu_2}\dots \delta_{\mu_{n-1}\mu_n} + \textit{permutations}\right)
\end{eqnarray}

However, the contraction over any pair of indices yields zeroes because $\frac{1}{|\vec r|}$ satisfies the Laplace equation.
Therefore $C =0$.

The number $d N$ of the vortex structures on the  large sphere would be estimated as
\begin{eqnarray}
  d N = 4 \pi \rho(R) R^2 d R
\end{eqnarray}
where $\rho(R)$ is the distribution of distances between the vortex structures.

After some tensor algebra and symbolic angular integration \cite{MB8} we found the formula for $\sigma$ with separated averaging over the unit vector on a sphere $S_2$ and the random tensor $W$
\begin{eqnarray}
  && 5 \sigma^2 = \frac{9 }{2 \pi ^2}\VEV{\Omega_{\alpha\beta}^2}_W 
  4 \pi \int d R \frac{\rho(R)}{R^6};
\end{eqnarray}

This distribution is normalized as
\begin{eqnarray}
  4 \pi \int d R \rho(R) R^2  =1
\end{eqnarray}
Therefore, our expression involves a mean value of $1/R^8$
\begin{eqnarray}
  &&\VEV{\frac{1}{R^8}}  = \frac{\int d R \rho(R) R^{-6}} {\int d R\rho(R)  R^2 };\\
  &&\int d R \rho(R) R^{-6} = \frac{1}{4 \pi} \VEV{\frac{1}{R^8}} 
\end{eqnarray}

After that, we relate the variance to the mean squared vorticity of each vortex structure and the relative distance distribution of these tubes.
\begin{eqnarray}
  \sigma^2 = \frac{9 }{10 \pi ^2}\VEV{\Omega_{\alpha\beta}^2}_W \VEV{\frac{1}{R^8}} 
\end{eqnarray}

In our theory, with a cylindrical tube of size $L$
\begin{eqnarray}
&&\Omega_{\alpha\beta} = L\left(\delta_{\alpha z} I_\beta - \delta_{\beta z} I_\alpha\right);\\
&& I_z =0;\\
&& I_x + \i I_y  = \oint_C d \Gamma C = \nonumber\\
&&\frac{2 c R_0^3 }{3 (\mu -1)}\left[x^3 +\frac{3 \mu ^2 x^{1-2 \mu }}{1-2 \mu }-\frac{3 i |x|^{2-\mu }}{\mu -2}-i \mu  |x|^{-3 \mu }\right]_{x= -x_{max}}^{x = x_{max}} 
\end{eqnarray}

In the turbulent limit $ x_{max} \ra  \frac{P}{4 R_0}$ we find
\begin{eqnarray}
&& I_x + \i I_y \ra \frac{ c P^3  }{48(\mu -1)}
\end{eqnarray}
After integrating over the eigenvalues $a,b,c$, assuming perimeter $P$ fixed, we find the following expression for the variance of strain
\begin{eqnarray}
  &&\sigma^2 = \sigma^2  \frac{ 1}{3(16\pi)^2}\VEV{ L^2 P^6} \VEV{\frac{1}{R^8}};\\
  && \bar R = \VEV{\frac{1}{R^8}}^{-\frac{1}{8}} 
\end{eqnarray}

The variance cancels here and this brings us to the final result for the mean energy dissipation
\begin{eqnarray}\label{MeanField}
&&\VEV{\mathcal E} \propto    \VEV{ L P^3} \sqrt{\nu\sigma^5};\\
&&  \VEV{L^2 P^6} = 3 (16\pi)^2 \bar R^8 ;
\end{eqnarray}

Let us be specific about the geometry scale here: we choose the mean distance between vortex structures $\bar{R}$ as a universal length scale. The sizes of the individual \CVS{} surfaces vary and fluctuate, but this $\bar R$ is a global parameter of our system so that we can use it as a unit of length.

The energy dissipation (for a single \CVS{} surface) in this case scales as
\begin{eqnarray}
\VEV{\mathcal E} \propto   \bar R^4 \sqrt{\nu\sigma^5};
\end{eqnarray}

Note that the tube length $L$ and the unknown perimeter $P$ dropped from this relation.

Thus, we get a scaling relation in the turbulent limit, the same we assumed in previous papers \cite{M21c,M20c}
\begin{eqnarray}
     \sigma   \sim \left(\frac{\mathcal E }{\bar R^4 }\right)^{\frac{2}{5}} \nu^{-\frac{1}{5}} 
\end{eqnarray}

One could estimate the length $L$ of the tube from the requirement that the tube volume becomes $V \sim \bar R^3$.
Such a volume in our system is likely to be occupied by at least one more vortex structure. In this case, there is a collision terminating the long tube.

Using the estimate \eqref{Area} of the area we find the estimate for $L$
\begin{eqnarray}
L  \sim \frac{\bar R^3}{\mbox{Area}} \propto R^3 P^{-1} h^{-1}
\end{eqnarray}

Combined with the above estimate $L P^3 \propto \bar R^4$ we find the new scaling relations
\begin{eqnarray}\label{PLEstimate}
&&\frac{\bar R}{P} \propto \bar R^{\oh} h^{-\oh} \propto  \nu^{-\frac{3}{10}};\\
&&\frac{L}{R} \propto \bar R^{\frac{3}{2}} h^{-\frac{3}{2}} \propto  \nu^{-\frac{9}{10}}\\
&& \frac{L}{\sqrt{\mbox{Area}}} \propto L^{\frac{3}{2}}\bar R^{-\frac{3}{2}}\propto \nu^{-\frac{27}{20}} ;\\
&&\frac{L}{P} \propto \bar R^{2} h^{-2} \propto   \nu^{-\frac{6}{5}} 
\end{eqnarray}

These estimates vindicate the assumption of the dilute gas approximation: the mean perimeter $P$ of our vortex particles is much smaller than their mean separation $\bar{R}$. 

At the same time, the perimeter $P$ of the structure cross-section is much smaller than its length, justifying the cylindrical approximation.
Finally, the perimeter $P$ is much larger than the thickness $h$, justifying the approximation of an infinitely thin vortex surface in the turbulent limit.

\section{"multifractals"}

Our hyperbolic vortex tube $ |y| |x|^{\mu} = \mbox{const} $ is smooth, but its shape has a random power index $\mu$.
The fluctuating power indexes imitate the multifractal scaling laws without any conventional fractals present.

Velocity field in the \CVS{} theory is potential everywhere in space, except the thin layers surrounding our hyperbolic vortex sheets.

The velocity difference between two neighboring points $\vec r_1, \vec r_2$  in the potential flow with scale as $\vec r_1 - \vec r_2$ as the harmonic potential has no singularity. 

However, with some probability, these two points will be separated by a vortex surface, in which case there will be a finite gap $\Delta \vec v$.

The estimate of the velocity difference moments would involve integration over the translation, rotation, and eigenvalues of the strain tensor.

Schematically, assuming $\vec r_1 = (x_1,y_1, 0)$ outside our vortex tube, and $\vec r_2 =(x_2,y_2,z)$ inside, we have (with $\hat r$ being unit vector of $\Delta \vec r$, $\eta_1 = x_1 + \i y_1, \eta_2 = x_2 + \i y_2$):
\begin{eqnarray}\label{MomentsEstimate}
&&\VEV{(\Delta \vec v\cdot \hat r)^n} \sim \int d P_{\sigma}(W)\int_{\Sigma(\Omega)} d x_1 d y_1 d x_2 d y_2 d z \nonumber\\
&&\left(\Re \left(a x_1 - \i b y_1 + F_+(\eta_1) + a x_2 + (a-c) \i y_2 \right) \frac{\eta_2^*-\eta_1^*}{|\eta_2-\eta_1|}\right)^n;\\
&& \Sigma(\Omega): |y_{1}||x_{1}|^{\mu}>1,  |y_{2}||x_{2}|^{\mu}<1,\Omega\cdot (\vec r_2- \vec r_1) = (|\Delta \vec r|,0,0) ;\\
&& \eta_1 = x_1 + \i y_1, \eta_2 = x_2 + \i y_2;
\end{eqnarray}

The measure for the traceless random $3\times3$ matrices $d P_{\sigma}(W) $ is presented in  \eqref{Pab}. The nontrivial part is the $SO(3)$ integration measure over the $3D$ rotations $\Omega$ involved in that measure. The distribution of the ratio of eigenvalues $a,b,c$ in \eqref{Pab} leads to the distribution of the index $\mu=1-\frac{c}{a}$.

\begin{figure}%
    \centering
     \includegraphics[width=\textwidth]{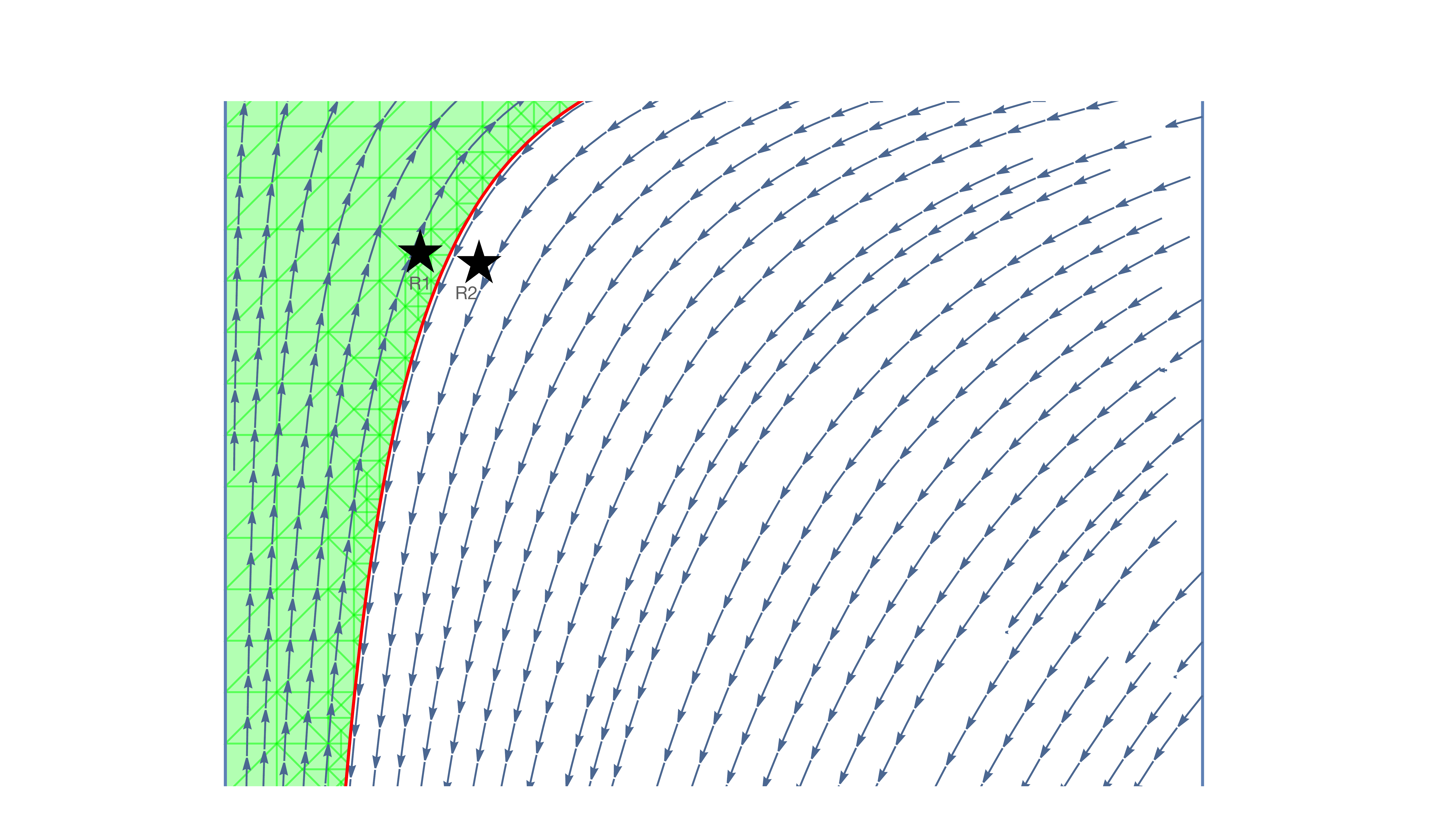}
    \caption{The two points measuring velocity difference are pinching the vortex surface (red)}.
    \label{fig::FlowWithDiff}
\end{figure}

Integrating over the highest eigenvalue $c$ at fixed $\mu $ we get \cite{MB3}:
\begin{eqnarray}
&&\VEV{(\Delta \vec v\cdot \hat r)^n} \sim\int d \Omega \int_{\Sigma(\Omega)} d x_1 d y_1 d x_2 d y_2 d z \int_{\frac{3}{2}}^3 d \mu Q_n(\mu) \nonumber\\
&&\left(\Im  \left(\frac{2 }{w_1(\eta_1)^{\mu }} + \frac{\mu  (y_1-y_2)+\i (x_1+x_2+2 \i y_1)}{\mu -1}\right) \frac{\eta_2^*-\eta_1^*}{|\eta_2-\eta_1|}\right)^n  ;\\
&& \Sigma(\Omega): \Omega\cdot (\vec r_2- \vec r_1) = (|\Delta \vec r|,0,0) ;\\
&&Q_n(\mu)  = \frac{\sqrt{3}\Gamma \left(\frac{n+5}{2}\right)(3-\mu) \mu  (2 \mu -3) (\mu -1)^n}{2\sqrt{\pi}((\mu -3) \mu +3)^{\frac{1}{2} (n+5)}  };\\
&& \eta_1 = x_1 + \i y_1, \eta_2 = x_2 + \i y_2;\\
&& \eta_1 = w_1 - \i w_1^{-\mu}
\end{eqnarray}
In this integral, the pair of points $(x_1,y_1, z), (x_2,y_2,0)$ are sliding along the curve while staying on different sides, separated by a fixed distance $\Delta r$. (see Fig.\ref{fig::FlowWithDiff}. 

The computation of this multi-dimensional integral is a whole new numerical problem, calling for supercomputer resources. As a rough estimate, demonstrating the phenomena of nontrivial effective dimension $\zeta(n, \log \Delta r)$ we studied the following model version of this integral, with the saddle point evaluation with parameter $x$ related to $\Delta r$
as follows
\begin{eqnarray}
&&\VEV{|\Delta V|^n}_{model} = \mbox{const} \int_{\frac{3}{2}}^3 d \mu Q_n(\mu) \left(x^2 + \mu^2 x^{-2 \mu}\right)^{\frac{n}{2}};\\
&& x: \Delta r = \mbox{const}\sqrt{ x^2 +  x^{-2 \mu}};
\end{eqnarray}

This integral is estimated by a saddle point equations
\begin{eqnarray}
&&\frac{Q_n'(\mu)}{Q_n(\mu)} + \frac{n}{2} \pp{\mu} \log \left(x^2 + \mu^2 x^{-2 \mu}\right) =0;\\
&& \log  \Delta r = \oh \log \left( x^2 +  x^{-2 \mu}\right);\\
&& \zeta(n ,\log  \Delta r ) = \pbyp{\log \VEV{|\Delta V|^n}}{\log \Delta r}   =\nonumber\\
&&n \frac{\pp{x} \log \left(x^2 + \mu^2 x^{-2 \mu}\right)}{\pp{x} \log \left(x^2 +  x^{-2 \mu}\right)}
\end{eqnarray}

\begin{figure}%
    \centering
      \includegraphics[width=\textwidth]{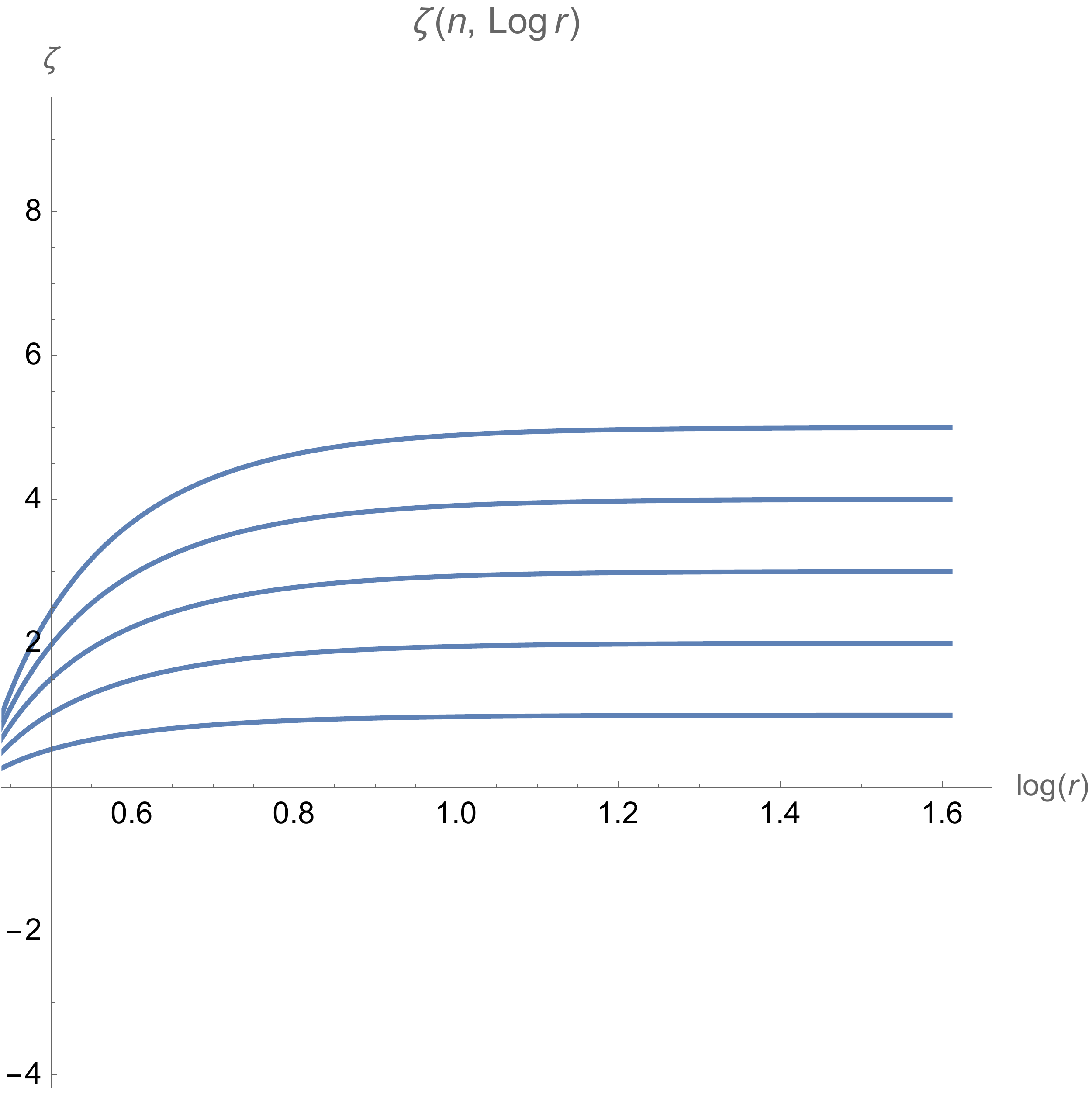}
    \caption{Effective fractal dimension $\zeta(n, \log  \Delta r)$ as a function of $\log \Delta r$ for $n =1\dots,5$ in a model approximation to the integral \eqref{MomentsEstimate}. The curves for $\zeta(n, \log \Delta r)$ approach $n$ at 
    $\log  \Delta r \ra \infty$.}
    \label{fig::Zeta}
\end{figure}

Solving these parametric equations numerically and eliminating $\mu, x$, we compute the dependence of $\zeta(n ,\log  \Delta r ) $ of its two variables.
We find the following plot Fig . \ref{fig::Zeta}.

This plot is very far from the observed values in DNS \cite{S19} where it is close to K41 straight line  $n/3$ reaching a plateau at $n \sim 12$. Either this approximation is too crude, or the observed turbulence corresponds to a different \CVS{} solution, without cylindrical geometry. The last possibility seems most likely.

In a recent work of Sreenivasan and Yakhot \cite{SY21} it was argued that with the "renormalized" Bernoulli formula for the pressure
\begin{eqnarray}
&&p = -\oh \alpha\vec v^2;\\
&& \alpha \approx 0.95 
\end{eqnarray}
the Hopf chain of equations for the velocity moments $ \VEV{|\Delta v(\vec r)|^n}$ can be closed, providing relations for the multifractal exponents, closely matching those observed  in numerical simulations.

As we know that the Bernoulli equation is not valid in a turbulent flow, this observation looks like a paradox.

However, in the \CVS{} theory, these estimates make sense.

For the (majority of) points in space outside the vortex surface, the original Bernoulli formula, with $\alpha = 1$, holds.
We reinterpret coefficient $\alpha$  as a probability of having no vorticity in a given point, involved in the computation of the Hopf moments.

The unconditional probability of avoiding a randomly translated and rotated smooth surface at a given point in space equals $1$. However, there are several reasons for this probability $\alpha$ to be less than 1.

One is the finite thickness of the vortex sheet:
\begin{eqnarray}
h \propto \sqrt{\frac{\nu}{\sigma}} \sim \nu^{\frac{3}{5}} \sim \R^{-\oh}
\end{eqnarray}

With $\R \sim 10^4$, this thickness measures in percents of the mean radius, which by itself can explain the $5\%$ deviation from $1$.

Another reason is that in the conditions used to compute the Hopf equations for the velocity moments, a vortex surface must be present between the two points $\vec r_1, \vec r_2$. However, each of these two points belongs to the potential flow region.

The conditional probability of having a vortex sheet between two points would be finite, providing another source of renormalization of the Bernoulli law.

The arguments involved in \cite{SY21} were relying on the turbulent viscosity notion, which we do not know how to justify in our theory. Still, the turbulent viscosity is an existing effect in the complementary approach with fluctuating velocity field. Perhaps, our effect of avoiding the vortex sheet is a dual description of the same phenomenon of turbulent viscosity.

The complete theory would have to compute the moments from the first principle. Then we will find the full function of two variables  $n, \log |\Delta r|$, currently approximated by the multifractal scaling law.

\section{Wilson Loop Statistics}

Let us assume that we have some \CVS{} surface parameterized by the random tensor $W= \diag{a,b,c}$. 

There are space translations, factored by a cylindrical translation in the $z$ axis and parameters of the symmetric traceless matrix $W$: two eigenvalues $  c>0, a \in (-2 c, -c/2) $ plus arbitrary rotations $\Omega \in S_2$ of the coordinate system. 

As we described in the previous sections, we treat this $W$ as the background uniform strain created by other vortex structures far away from the surface's axis. Thus, one would think that this strain, adding up strains from a large number of randomly positioned vortex structures like this one, would be a random traceless symmetric tensor.

The $z$ axis of the tube automatically aligns with the leading eigenvector of $W$, the one with the largest eigenvalue $c>0$. The \CVS{} stability requires this.

There remains an integral over zero modes: translation and rotation of the coordinate system plus two eigenvalues of the random constant strain at infinity.  We should integrate over these parameters the observables computed for a particular solution of \CVS{}.

To be specific, let us consider the loop average
\begin{eqnarray}
&& W_C(\gamma) \propto  \int d^3 r_0 d \Omega d P(p,q) \exp{\i \gamma \oint_C \val(\vec r) d \ral}
\end{eqnarray}

The velocity circulation in the exponential comes only from the intersection points of the loop $C$ with our cylindrical surface. 
A simple loop intersects the surface in two points or does not intersect at all, so we have
\begin{eqnarray}
&&\Gamma_{1 2} = \oint_C \val d \ral =  \Gamma(\eta_2) - \Gamma(\eta_1)
\end{eqnarray}
Thus, up to extra terms independent of $\gamma$
\begin{eqnarray}
W_C(\gamma) \propto 
\int d^3 r_0 d \Omega d P(a,b)\Pi_S(\vec r_0,\Omega, \gamma)
\end{eqnarray}

The factor
\begin{eqnarray}
&&\Pi_S(\vec r_0, \Omega,\gamma) = \int_0^{1} d l_1 | C'(l_1)|\int_0^{1}  d l_2| C'(l_2)|\nonumber\\
&&\int_{\vec r_1 \in S} \int_{\vec r_2 \in S}  
\delta^3\left( \vec r_1 + \Omega \cdot(\vec r_0 - \vec C(l_1))\right)\delta^3\left( \vec r_2 + \Omega \cdot(\vec r_0 - \vec C(l_2))\right)\nonumber\\
&& \exp{\i \gamma (\Gamma(\eta_2) - \Gamma(\eta_1))}
\end{eqnarray}
was already computed in \cite{M21b}
\begin{eqnarray}
&&\Pi_S(\vec r_0,\Omega,\gamma) = \frac{|\vec \sigma_1|| C'(l_1)|}{\left|\vec \sigma_1 \cdot \Omega\cdot\vec C'(l_1)\right|} \frac{|\vec \sigma_2|| C'(l_2)|}{\left|\vec \sigma_2\cdot \Omega \cdot \vec C'(l_2)\right|}\theta(\vec r_1\in S)\theta(\vec r_2 \in S)\nonumber\\
&& \exp{\i \gamma (\Gamma(\eta_2) - \Gamma(\eta_1))};\\
&& \vec \sigma_1 = \vec \sigma(\vec r_1); \vec \sigma_2 = \vec \sigma(\vec r_2);\\
&& \vec r_1 = \Omega\cdot (\vec C(l_1) - \vec r_0);\vec r_2 = \Omega\cdot (\vec C(l_2) - \vec r_0);
\end{eqnarray}

This integral depends on the geometry of the problem: the relation between the loop $C$ and the cylinder.  The phase factor depends on the random strain parameters via known $\Gamma(\eta)$.

We are left with an integral over $2 d$ translations $ \vec r_0$ against the cylinder's axis and the $O_3$ rotations $\Omega$.

The two theta functions, restricting the points to be on the intersection of the surface and the loop, are zero if the point $\vec r_0$ moves from the surface farther than the maximal size of the loop $C$.

The integration region for $\vec r_0$ is some layer around the vortex surface, with the width equal to the maximal distance between two points on the loop.

The integration over rotations $\Omega$ is compact.
Therefore, this integral looks like a relatively straightforward integral from a computational point of view. 

As we already mentioned, there is no time-reversal symmetry that would reflect $\gamma$. The circulation is proportional to $a$ which is distributed in negative interval $(-2c, -\oh c)$.

As a consequence, our Wilson loop will be a complex number. Unfortunately, nobody has measured it in DNS, but its Fourier transform (over circulation, not over space coordinates) represents the circulation PDF measured with high accuracy.

This Fourier transform would simply mean replacement
\begin{equation}
    \exp{\i \gamma (\Gamma(\eta_2) - \Gamma(\eta_1)) } \Ra \delta\left((\Gamma(\eta_2) - \Gamma(\eta_1)) - \oint_C \vec v \cdot d \vec r\right)
\end{equation}

This delta function is welcome, as it will reduce the dimension of the remaining integration over $\vec r_0$.

After that, \Mathematica can compute this finite integral with arbitrary precision.

Another possibility is to compute the moments  of the circulation  $ M_n = \VEV{(\oint_C \vec v \cdot d \vec r)^n}$ which would correspond to the replacement
\begin{equation}
    \exp{\i \gamma (\Gamma(\eta_2) - \Gamma(\eta_1)) } \Ra \left((\Gamma(\eta_2) - \Gamma(\eta_1))\right)^n
\end{equation}

Thus, there is a steep but clear path to compute a Wilson loop using the above formulas.

\section{Discussion}
The \CVS{} theory offers an approach to turbulence that one can study analytically and numerically. 

Let us remind and discuss the general ideas behind this theory.

We are trying to solve the \NS{} equations for an infinite time evolution, given that the laminar flow is unstable. 
This unstable solution could cover the whole phase space of the fluid mechanics; however, we suspect that there are some low-dimensional attractors in that space.
The solution would cover the subspace of these attractors like the Newtonian dynamics covers the energy surface in ergodic motion.

We also know the phenomenon of anomalous dissipation: the dissipation in extreme turbulence occurs in the regions of high vorticity, which compensates for the viscosity factor in front of the enstrophy integral.

Furthermore, we conjecture that these regions of high vorticity are vortex surfaces. There are several reasons to believe so, both theoretical and (numerical or real) experimental sides.

The Hamiltonian dynamics of vortex surfaces \cite{M88,AM89} can be solved exactly in some important cases, and the resulting solutions have all nice properties, including anomalous dissipation.

We have found in our previous work \cite{M20c, M21b} that the steady-state of the vortex surface is a promising candidate for such an attractor in phase space. At first sight, it appears that these steady vortex surfaces have some local degrees of freedom, corresponding to the arbitrary shape of the surface, as long as the velocity gap satisfies certain integral equations, following from the minimization of the Hamiltonian.

Initially, we speculated that these surface degrees of freedom were described by some version of a solvable string theory \cite{M21b}. There are two kinds of surface degrees of freedom: the surface shape and the internal metric on this surface, described by the Liouville field.  

In fluid dynamics terms, the fluctuations of this internal metric correspond to the tangent motions of the fluid, equivalent to the reparametrization of this surface.

The shape of the surface was assumed frozen in the turbulent flow in \cite{M21b}, and the remaining internal metric was assumed to be responsible for the multifractal scaling laws for the moments of velocity circulation.

The new recent understanding eliminated the shape of the surface as degrees of freedom while placing no restrictions on the internal metric on this surface.

Some surfaces are "more equal" than others because they are stable at the microscopic level.
These are the \CVS{} surfaces satisfying the stability equations \eqref{CVS}. 

It is truly remarkable that such a simple equation, involving just three letters, leads to so many specific consequences for the turbulent flow.

The negative normal strain presses vorticity into the \CVS{} surface from both sides, leading to a narrow Gaussian profile of vorticity in the normal direction. The width of this boundary layer goes to zero as some negative power of Reynolds number in the turbulent limit.

We assume that our flow and the surface are subject to a background strain, a random traceless matrix adding up from \BS{} tails of many other vortex structures located far away from this one. Such random strain is described by eigenvalues $a,b,c$ adding up to zero. The parameters $a,b,c$ obey a Gaussian distribution multiplied by a famous Vandermonde determinant.

The normal strain is uniform all around our vortex surface  $S_{n n} = -c$, so that we have one of the stability criteria fulfilled if the axis of the vortex tube points along the main eigenvector of the strain, with a positive eigenvalue $c$.
The tangent strain is degenerate: there is zero strain along the tangent velocity gap. 

The energy dissipation occurs at the surface, and we have demonstrated \cite{M21c} that this surface integral is conserved in the Navier-Stokes dynamics in the turbulent limit.

Let us stop and think about this conservation law. It is \textit{not} an Euler conservation law, violated by viscosity, like the energy conservation. Instead, we have found a previously unknown \NS{} conservation law, and it comes about as a result of the exact cancellation of the advection, vortex stretching, and diffusion terms in the \NS{} equation used for the time derivative of the enstrophy.

The Euler dynamics would lead to an explosion or decay of enstrophy were it not offset by the diffusion in full \NS{} equation. For any other vortex surface but the \CVS{} the diffusion would not cancel the vortex stretching so that the notorious energy dissipation constant $\mathcal E$ would not be constant.

The Kolmogorov theory kept this paradox unresolved for 80 years. The basic parameter of the theory was constant with one definition (three-point function of velocity), but not with an alternative one (enstrophy), and nobody knew why.

As we see it now, $\mathcal E $ is conserved because of the \CVS{} stability of the vortex surfaces hidden behind the rough picture of the energy flow in Fourier space. One cannot describe the shape of the discontinuity surface by a finite (or even convergent) set of Fourier coefficients of the velocity field.

We describe the turbulence phenomenon as a fluctuating geometry instead of fluctuating velocity field. In the modern QFT, such a pair of complementary descriptions represents Duality.

In the ADS-CFT theory \cite{ADS} the conformal field theory (fluctuating vector gauge field) is dual to the fluctuating geometry of curved space. The strong coupling of the conformal field theory corresponds to the (solvable) weak coupling limit of the fluctuating geometry.

A more direct analogy is the description of the Burgers' one-dimensional turbulence of compressible fluid as a statistics of shocks- one-dimensional analogs of our vortex surfaces. This shock statistics is also a dual theory of geometric objects, weakly fluctuating in the extreme turbulent limit when they reduce to step functions of the coordinate.

In our theory, the energy flows from large scales to small scales. It dissipates at small scales in thin boundary layers of the \CVS{} surface, but this spatial picture is different from the hierarchical cascade of energy "from a big eddy to a smaller one." 

The only way to reconcile these two pictures is to assume the Russian doll of nested tubes, passing the energy from the outer tube to the inner ones, with each tube dissipating energy as discussed above. Someone still has to find such a solution of the \CVS{} equations.

The conservation of energy dissipation on the CVS surface makes this parameter $\mathcal E$ an appropriate scaling factor for the K41 theory viewed as dimensional analysis, which describes the medium Reynolds numbers so well.

We hope that there exist full 3D solutions to the \CVS{} equations, with the finite area and finite volume inside. If found, such solutions would be the best candidates for turbulent statistics. Lacking that, we can analyze the cylindrical solutions as some approximation to reality or proof of concept.

Let us come back to our cylindrical solution.

This solution describes a vortex tube much larger than the viscous scale. All vorticity collapses to the boundary layer; this layer's width will decrease as a positive viscosity power in the turbulent limit. 

The flow is, therefore, potential both inside and outside.
The normal velocity vanishes on both sides of the surface of the tube, but the tangent velocity is finite, and there is a tangent gap.

Hollow vortex tubes were observed in the recent DNS \cite{Tubes19}, but it is hard to tell until some research would specifically hunt for hollow tubes in turbulent flow simulations.

The uniform strain tensor, distributed as a Gaussian random traceless matrix, is forcing our flow.

The origin of this random force is the accumulation of a large number of $1/R^3$ tails of the Biot-Savart laws for other remote vortex tubes with random locations and random orientations.

The stability condition $ S_{n n} <0$, coming from the microscopic analysis of the \NS{} equation in a boundary layer, fixes the sign of the flow. 

The same requirement follows from the \KH{} stability of the vortex surface against the low wave-length perturbations. The high wave-length perturbations of the sharp vortex surface are always unstable, but presumably, viscous diffusion stabilizes those perturbations as it happens in the \BT{} solution.

This stability condition is the origin of the flow's irreversibility. We choose the parameters so that the energy dissipation (and equal energy pumping) stays finite in the limit of vanishing viscosity.

We considered these closed vortex surfaces (vortex tubes) as a dilute gas, with a large mean distance between tubes, compared to their size.

In this dilute gas approximation, we derived the Gaussian random strain from an infinite number of remote vortex structures, adding to the \BS{} formula for the strain at large distances and averaging to the random constant stress tensor due to the Central Limit theorem.

This calculation leads to a self-consistency relation \eqref{MeanField}  for the mean distance $\bar R$  between vortex tubes, their length $L$, and the perimeter $|C|$ of the cross-section loop.

The perimeter being much smaller than the mean distance between the vortex tubes vindicates the underlying assumption of an ideal gas of vortex tubes. 

The statistical distribution of the shape index $\mu$, following from the Gaussian distribution of the background strain, offers the microscopic explanation for the "multifractal" scaling laws for the moments of velocity difference.

The moments of velocity difference are a more complex function of $n, |\Delta r|$ than just  $|\Delta r|^{\zeta(n)}$, which could still be approximated by the power laws at specific integer values of $n$. 

Our estimate \eqref{MomentsEstimate} provides the mechanism for the multifractal law, although these multidimensional integrals require supercomputer resources to compute with precision. The rough estimate we made in this paper demonstrates the imitation of the multifractal, though the effective index $\zeta(n,\log \Delta r)$ in this estimate is not even close to the experiment.

The SY phenomenological theory \cite{SY21} relies on the postulate of the renormalized Bernoulli law for local values of pressure in the turbulent flow. We interpret this renormalized law $p = -\oh \alpha \vec v^2$ as a probability $\alpha$ to have no vorticity in the point under study times the local pressure in a potential flow. 

The velocity differences in the piecewise potential flow can be singular: the vortex tube's surface separates the pair of points with a certain probability. 

In this case, the pressure obeys the Bernoulli formula on each side. The velocity difference is dominated by the tangent velocity gap, scaling as some power of the size of the \CVS{} surface. 

This topological observation resolves the paradox of singular velocity moments in the potential flow of \cite{SY21}.
With this interpretation, the multifractal law is not exact -- it is only an approximation to the future microscopic theory.

Using the dilute gas approximation, we computed the PDF for the energy dissipation normalized by the third power of the perimeter of the loop. It is plotted in log-log scale at \ref{fig::Dissipation}. This curve is universal in our solution; there are no model approximations nor any dimensionless parameters.

In the same way, one can compute the distribution of the circulation and other observables, such as the Wilson loop $\VEV{\exp{\i \gamma \oint_C \val d \ral}}$.

\section{Summary}
\begin{itemize}
    \item \textit{Attractor} in the \NS{} equation is found: vortex tubes with singular shape, parametrized by the Gaussian random symmetric traceless matrix \eqref{Pab} in addition to random 3D location and random scale.
    \item \textit{Irreversibility} follows from the inequality on the local normal strain at the vortex tube: $S_{n n} <0$. This inequality is required for microscopic stability.
    \item \textit{Enstrophy} is conserved as a consequence of the stability equation: $\hat S \cdot \Delta \vec v=0$.
    \item \textit{Stochasticity} is explained by the accumulation of contributions to the background strain from a large number of uncorrelated remote vortex tubes via \BS{} law \eqref{BSStrain}. 
     \item \textit{Analytical Solution} is presented for the flow with cylindrical geometry \eqref{velocityField}. Far away from the tubes, there is a potential flow with random uniform background strain. 
    \item \textit{Multifractals} could be imitated by fluctuations of the shape of the \CVS{} solution. We found some multidimensional integrals describing the moments of velocity differences and velocity circulations. Numerical computation of these integrals would be a subject of a large numerical project outside our present scope.
    \item \textit{The rough estimate} of scaling laws for the moments of velocity difference for a cylindrical solution leads to a mismatch with the DNS, which leaves open the search for the full 3D solution of the \CVS{} equations.
\end{itemize}
\section*{Acknowledgments}

I am grateful to Theo Drivas, Dennis Sullivan, Katepalli Sreenivasan, Victor Yakhot, and the Stony Brook University Einstein seminar participants for stimulating discussions of this work. The help from Arthur Migdal with \Mathematica is also greatly appreciated.

I also appreciate thoughtful comments of my referee.
This work is supported by a Simons Foundation award ID $686282$ at NYU.

\appendix{
\section{The Cylindrical CVS equations}

Let us re-derive the cylindrical \CVS{} equations of \cite{M21c} without unnecessary assumption of real and equal functions $d \Gamma_\pm$  in \eqref{Vgamma}.

The normal vector in complex notation
\begin{eqnarray}
 \sigma =  \i C'(\theta)
\end{eqnarray}
The normal projection of velocity field
\begin{equation}
     v_x \sigma_x +  v_y \sigma_y = \Re (v_x - \i v_y) \sigma =
   \frac{\Im \left( (v_x - \i v_y) C'(\theta) \right)}{|C'(\theta)|}
\end{equation}

The first \CVS{} equation (vanishing normal velocity on each side) becomes
\begin{eqnarray}\label{Neumann}
   \Im\left( V_\pm(C(\theta)) C'(\theta)\right) =0;\forall \theta
\end{eqnarray}

The second and third \CVS{} equations require the computation of the strain related to the complex velocity $F_\pm = \phi'_\pm $. 
The strain on each side is a $3 \times 3$ matrix 
\begin{eqnarray}
  \hat S_\pm(\eta) = \left(
\begin{array}{ccc}
 \Re F'_\pm(\eta)+a & - \Im F'_\pm(\eta) & 0 \\
 -\Im F'_\pm(\eta) & - \Re F'_\pm(\eta) +b & 0\\
  0         &0                 &c\\
\end{array}
\right)
\end{eqnarray}

Let us introduce notations
\begin{eqnarray}
&&F(\eta) = \oh(F_+(\eta) + F_-(\eta));\\
&&\Delta F(\eta) = F_+(\eta) - F_-(\eta);\\
&&p = (a+b)/2;\\
&&q =(a-b)/2
\end{eqnarray}

The null vector equation $(\hat S_+(\eta) + \hat S_-(\eta)) \cdot \Delta \vec v =0$ provides the following complex equation
\begin{eqnarray}\label{eq1}
   && \Delta F^*(\eta)(F'(\eta) + q )  +  p  \Delta F(\eta) =0;\forall \eta \in C ;
\end{eqnarray}

The product 
\begin{eqnarray}
\Gamma(\theta) = C'(\theta) \Delta F(C(\theta))
\end{eqnarray}
is real, in virtue of the Neumann boundary conditions.
The difference of the two Neumann equations \eqref{Neumann} reduces to $\Im \Gamma =0$.

Thus, multiplying \eqref{eq1} by $C'(\theta)C'^*(\theta)$ and using the fact that $\Gamma(\theta)$ is real  we find
\begin{eqnarray}\label{NullVector}
   && C'(\theta) \left(F'(C(\theta)) + q \right)  +  p  C'^*(\theta) =0;\forall \eta \in C ;
\end{eqnarray}

This equation is simpler than it looks: it is reduced to the total derivative.
\begin{eqnarray}
   \dd{\theta} \left( F(C(\theta))+ q C(\theta) + p C^*(\theta) \right) =0 
\end{eqnarray}

The generic solution is
\begin{eqnarray}\label{Ceq}
    F(C(\theta))+ q C(\theta) + p C^*(\theta) = A
\end{eqnarray}
with some complex constant $A$. 

Plugging it back to the \eqref{Neumann} we have everything cancel except one term
\begin{eqnarray}
   \Im\left(A C'(\theta)\right) =0;\forall \eta =C(\xi)
\end{eqnarray}

The  nontrivial solution for $C(\theta)$ would correspond to $A =0$.

This solution is equivalent to the equation \eqref{CVSCyl} we stated in the text.
\section{Symmetric Traceless Random Matrices}
 
 Let us investigate the general properties of the measure \eqref{GaussTensor}. First, the trace of the matrix is invariant with respect to orthogonal transformations
 \begin{eqnarray}
 &&W \Ra O^T \cdot W \cdot O;\\
 && O^T = O^{-1};\\
 && \tr W \Ra \tr W
 \end{eqnarray}
 therefore this measure stays $O(n)$ invariant after insertion of the delta function of the trace.
 
 The mean value of  $\tr W^2$ can be computed by the following method. Consider the normalization integral
 \begin{eqnarray}
 Z_n(\sigma) = \int d P_\sigma(W)
 \end{eqnarray}
 By rescaling the matrix elements $W_{i j} = \sigma w_{i j}$ we find the property
 \begin{eqnarray}\label{Zint}
 Z_n(\sigma) = \sigma^{\frac{(n+2)(n-1)}{2}} Z_n(1)
 \end{eqnarray}
 
 Taking the logarithmic derivative of the original integral we get the identity
 \begin{eqnarray}
 \frac{\sigma Z_n'(\sigma)}{Z_n(\sigma)} = \frac{\int d P_\sigma(W) \tr W^2}{\sigma^2 Z_n(\sigma)} = \frac{\VEV{\tr W^2}}{\sigma^2}
 \end{eqnarray}

 On the other hand, taking the logarithmic derivative from \eqref{Zint} we find 
 \begin{eqnarray}
 \frac{\sigma Z_n'(\sigma)}{Z_n(\sigma)} = \frac{(n+2)(n-1)}{2}
 \end{eqnarray}
 This formula produces the trace relation \eqref{TraceRel} in the text.
 
 Higher moments of various products of matrix elements are also calculable analytically. We are studying these moments, and the generation function in \cite{MB3}.
}
 \bibliography{bibliography}

\begin{thebibliography}{10}
\expandafter\ifx\csname urlstyle\endcsname\relax
  \providecommand{\doi}[1]{doi:\discretionary{}{}{}#1}\else
  \providecommand{\doi}{doi:\discretionary{}{}{}\begingroup
  \urlstyle{rm}\Url}\fi

\bibitem{VTubes}
T.~Ishihara, T.~Gotoh and Y.~Kaneda, {\em Annual Review of Fluid Mechanics}
  {\bf 41}, 165 (12 2008), \doi{10.1146/annurev.fluid.010908.165203}.

\bibitem{Tubes19}
D.~Buaria, A.~Pumir, E.~Bodenschatz and P.~K. Yeung, {\em New Journal of
  Physics} {\bf 21},   043004 (Apr 2019), \doi{10.1088/1367-2630/ab0756}.

\bibitem{S19}
K.~P. Iyer, K.~R. Sreenivasan and P.~K. Yeung, {\em Phys. Rev. X} {\bf 9},
  041006 (Oct 2019), \doi{10.1103/PhysRevX.9.041006}.

\bibitem{M19d}
A.~Migdal, Analytic and numerical study of navier-stokes loop equation in
  turbulence  (2019), \href{http://arxiv.org/abs/arXiv:1908.01422v1}{{\ttfamily
  arXiv:1908.01422v1}}.

\bibitem{M93}
A.~Migdal, Loop equation and area law in turbulence, in {\em Quantum Field
  Theory and String Theory\/},  eds. L.~Baulieu, V.~Dotsenko, V.~Kazakov and
  P.~Windey (Springer {US}, 1995) pp. 193--231.

\bibitem{VortexGasCirculation20}
G.~Apolinario, L.~Moriconi, R.~Pereira and V.~valadão, {\em PHYSICAL REVIEW E}
  {\bf 102},   041102 (10 2020), \doi{10.1103/PhysRevE.102.041102}.

\bibitem{QuantumCirculation21}
N.~P. M\"uller, J.~I. Polanco and G.~Krstulovic, {\em Phys. Rev. X} {\bf 11},
  011053 (Mar 2021), \doi{10.1103/PhysRevX.11.011053}.

\bibitem{BURGERS1948}
J.~Burgers, A mathematical model illustrating the theory of turbulence, in {\em
  Advances in Applied Mechanics\/},  eds. R.~{Von Mises} and T.~{Von Kármán}
  (Elsevier, 1948) pp. 171 -- 199.

\bibitem{TW51}
A.~A. {Townsend}, {\em Proc. Roy. Soc. Lond. Ser. A.} {\bf 208(1095)},
  534–542  (1951), \doi{10.1098/rspa.1951.0179.}

\bibitem{M21c}
A.~Migdal, Confined vortex surface and irreversibility. 1. properties of exact
  solution  (2021), \href{http://arxiv.org/abs/2103.02065v10}{{\ttfamily
  arXiv:2103.02065v10 [physics.flu-dyn]}}.

\bibitem{M21a}
A.~Migdal, {\em Physics of Fluids} {\bf 33},   035127  (2021),
  \href{http://arxiv.org/abs/https://doi.org/10.1063/5.0044724}{{\ttfamily
  https://doi.org/10.1063/5.0044724}}, \doi{10.1063/5.0044724}.

\bibitem{KS21}
K.~Shariff and G.~E. Elsinga, {\em Physics of Fluids} {\bf 33},   033611
  (2021),
  \href{http://arxiv.org/abs/https://doi.org/10.1063/5.0045243}{{\ttfamily
  https://doi.org/10.1063/5.0045243}}, \doi{10.1063/5.0045243}.

\bibitem{MB16}
A.~Migdal, { General vortex sheet}
  https://www.wolframcloud.com/obj/sasha.migdal/Published/GeneralVortexSheet.nb
  (August, 2021).

\bibitem{DL21}
Wikipedia, {Double layer (surface science)} --- {W}ikipedia{,} the free
  encyclopedia  (2021), [Online; accessed 24-January-2021].

\bibitem{MB14}
A.~Migdal, { Algebraic cvs}
  https://www.wolframcloud.com/obj/sasha.migdal/Published/AlgebraicCVS.nb
  (August, 2021).

\bibitem{MB15}
A.~Migdal, { Hyperbolic flow}
  https://www.wolframcloud.com/obj/sasha.migdal/Published/HyperbolicFlow.nb
  (August, 2021).

\bibitem{BrT}
Wikipedia, { {Brouwer Fixed-point Theorem}}
  \url{https://en.wikipedia.org/wiki/Brouwer_fixed-point_theorem},  (2021),
  [Online; accessed 11-June-2021].

\bibitem{M88}
A.~A. Migdal, { Random surfaces and turbulence}, in {\em Proceedings of the
  International Workshop on Plasma Theory and Nonlinear and Turbulent Processes
  in Physics, Kiev, April 1987\/},  ed. V.~G. Bar’yakhtar (World Scientific,
  1988), p. 460.

\bibitem{AM89}
M.~E. Agishtein and A.~A. Migdal, {\em Physica D: Nonlinear Phenomena} {\bf
  40}, 91   (1989), \doi{https://doi.org/10.1016/0167-2789(89)90029-8}.

\bibitem{RSM}
Wikipedia, { {Random Matrix}}
  \url{https://en.wikipedia.org/wiki/Random_matrix},  (2021), [Online; accessed
  30-September-2021].

\bibitem{MB3}
A.~Migdal, { Matrix integrals}
  https://www.wolframcloud.com/obj/sasha.migdal/Published/MatrixIntegrals.nb
  (May, 2021).

\bibitem{MB8}
A.~Migdal, { N dimensional tensor algebra and gradients}
  https://www.wolframcloud.com/obj/sasha.migdal/Published/NDimensionalTensorMath.nb
  (Mar, 2021).

\bibitem{M20c}
A.~Migdal, {\em International Journal of Modern Physics A} {\bf 35},   2030018
  (November 2020), \href{http://arxiv.org/abs/2007.12468v7}{{\ttfamily
  arXiv:2007.12468v7 [hep-th]}}, \doi{10.1142/s0217751x20300185}.

\bibitem{SY21}
K.~R. Sreenivasan and V.~Yakhot, Dynamics of three-dimensional turbulence from
  navier-stokes equations  (2021).

\bibitem{M21b}
A.~Migdal, {\em International Journal of Modern Physics A} {\bf 36},   2150062
  (2021),
  \href{http://arxiv.org/abs/https://doi.org/10.1142/S0217751X21500627}{{\ttfamily
  https://doi.org/10.1142/S0217751X21500627}}, \doi{10.1142/S0217751X21500627}.

\bibitem{ADS}
Wikipedia, { {AdS-CFT correspondence} --- {W}ikipedia{,} the free encyclopedia}
  \url{https://en.wikipedia.org/wiki/AdS/CFT_correspondence},  (2019), [Online;
  accessed 28-October-2021].

\end{thebibliography}

\section{Figures}
\newpage


\end{document}